\documentclass[useAMS,usenatbib]{mn2e}

\usepackage{graphicx}
\usepackage{deluxetable}
\usepackage{aalongtable}
\usepackage{epsfig}

\usepackage{txfonts}
\def\ha{{\rm H$_\alpha$\,\,}}
\def\near{$\sim$}

\newcommand{\bfig}{\begin{figure*}}
\newcommand{\efig}{\end{figure*}}
\newcommand{\btab}{\begin{table*}}
\newcommand{\etab}{\end{table*}}
\newcommand{\bcen}{\begin{center}}
\newcommand{\ecen}{\end{center}}

   \title{Exploring pre-main sequence variables of ONC: The new variables}

\author[Padmakar Parihar, et al.] {Padmakar Parihar$^{1}$\thanks{E-mail:
psp@iiap.res.in}, Sergio Messina$^{2}$\thanks{E-mail:sme@oact.inaf.it},
  Elisa Distefano$^{2}$\thanks{E-mail:eds@oact.inaf.it}, Shantikumar N.S. $^{1}$\thanks{E-mail:shanti@iiap.res.in}, and Biman J. Medhi$^{3}$ \\
$^{1}$Indian Institute of Astrophysics, Bangalore 560034, India\\
$^{2}$INAF-Catania Astrophysical Observatory, Italy\\
$^{3}$Aryabhatta Research Institute of Observational
    Sciences (ARIES), Manora Peak, Nainital -263129, India\\ }
\begin{document}

\date{Accepted ..... Received ......; in original form .....}
\pagerange{\pageref{firstpage}--\pageref{lastpage}} \pubyear{....}

\maketitle

\label{firstpage}

\begin{abstract}

Since 2004, we  have been engaged in a  long-term observing program to
monitor young  stellar objects in  the Orion Nebula Cluster.   We have
collected about two thousands frames in V,~R, and I broad-band filters
on  more than  two hundred  nights distributed  over  five consecutive
observing  seasons.  The  high-quality  and time-extended  photometric
data give  us an opportunity  to address various  phenomena associated
with young  stars.  The prime motivations  of this project  are \it i)
\rm to explore various  manifestations of stellar magnetic activity in
very  young low-mass stars;  \it ii)  \rm to  search for  new pre-main
sequence eclipsing binaries; and \it iii) \rm to look for any EXor and
FUor like  transient activities associated  with YSOs.  Since  this is
the first paper on this program, we give a detailed description of the
science drivers, the observation  and the data reduction strategies as
well. In  addition to  these, we  also present a  large number  of new
periodic variables  detected from our first five  years of time-series
photometric data.   Our study reveals that  about 72\% of  CTTS in our
FoV are  periodic, whereas,  the percentage of  periodic WTTS  is just
32\%.  This indicates that  inhomogeneities patterns on the surface of
CTTS of  the ONC stars  are much more  stable than on WTTS.   From our
multi-year monitoring  campaign we found that  the photometric surveys
based  on  single-season are  incapable  of  identifying all  periodic
variables.  And  any study on  evolution of angular momentum  based on
single-season surveys must be carried out with caution.

\end{abstract}

\section{Introduction}

The Orion  Nebula Cluster  (ONC) is an  excellent target  for studying
young stellar objects (YSOs).  It contains a few thousands of pre-main
sequence (PMS) stars within $\sim$15 arc-minutes (2 pc) of the central
Trapezium   stars  (Herbig   \&  Terndrup   1986;   Hillenbrand  1997;
Hillenbrand \&  Hartmann 1998). At  a distance, 450$\pm$70 pc,  ONC is
the  nearest high-mass  star-forming region  where one  can  find very
massive stars of  $\sim$25 solar mass ($\theta^1$ Ori  C) to sub-solar
objects having mass well below the hydrogen burning limit (Hillenbrand
1997;  Lucas \&  Roche 2000).   From recent  studies, it  appears that
formation of  stars in the ONC started  about 10 Myr back  and, in the
beginning, the formation  process was very slow.  Later,  due to large
scale contraction in the ONC parental cloud, the activity went through
a rapid phase of  star formation.  The mean age of the  ONC is about 1
Myr, which  characterizes an epoch  when large fraction of  stars were
born.   An age  spread  of 2  Myr around  this  mean age  is found  by
previous  researchers (Hillenbrand 1997;  Palla et  al. 2005;  Huff \&
Stahler 2006; Jeffries  2007).  The ONC is located in  front of a very
extended and fully opaque molecular cloud (A$_V$ peaks at 80-100 mag),
which makes the background  star contamination almost insignificant in
the  optical region.   This means  that  except for  a few  foreground
stars,  all visible  objects can  be  straight away  considered to  be
cluster members.  Very intense radiation pressure and stellar winds of
the massive stars  have cleared most of the dust  from the ONC (O'Dell
2001)  and that is  why about  80\% cluster  members are  subjected to
relatively low  visual extinction, A$_V$  ranging from 0.0 to  2.5 mag
(Hillenbrand 1997).  The low-mass ONC stars are very strong sources of
X ray emission and the  median luminosity of these objects is L$_X\sim
30.25$  erg sec$^{-1}$  (L$_X$/L$_{bol}$ $\sim$10$^{-3.5}$),  which is
three orders of magnitude more intense than the solar X-ray luminosity
(Flaccomio  et  al.  2003).   Recent  studies  reveal  that the  X-ray
luminosity  of low-mass  stars  increases with  the  stellar mass  and
decreases with the age, but it seems to be independent of the rotation
period (Flaccomio  et al.   2003; Stassun et  al.  2004;  Preibisch et
al. 2005).

  In  the past,  for more  than one  decade, the  ONC and  the regions
  close-by  (ONC flanking  fields) were  photometrically  monitored by
  various researchers  with varying degree  of sensitivity as  well as
  spatial coverage  (Mandel \& Herbst  1991; Attridge \&  Herbst 1992;
  Eaton et al. 1995; Choi \&  Herbst 1996; Stassun et al. 1999; Herbst
  et  al. 2000;  Carpenter et  al. 2001;  Rebull 2001;  Herbst  et al.
  2002).   Except  the  long-term  observing  program  of  Herbst  and
  collaborators  of Van  Vleck  Observatory (VVO)  and near  infra-red
  (NIR)  survey  carried  out  by  Carpenter  et  al.   (2001),  other
  monitoring programs were primarily focused on a single goal and that
  was to explore the evolution of angular momentum of the stars in the
  PMS phase, by determining the rotation periods from the light curves
  produced by modulation of light  due to hot/cool spots.  We realized
  that  moderate  size telescopes  equipped  with  wide field  imaging
  camera can be very effectively used to address a variety of valuable
  scientific problems  related to PMS  stars. And hence  using various
  observing facilities accessible to  our group, since January 2004 we
  have initiated  a long-term monitoring  program on the PMS  stars of
  the ONC.  Since this is the first paper on this project, we not only
  report the  results on the identification of  new periodic variables
  but also describe the science  drivers, the observation and the data
  reduction  in  some  more  detail.   In  Sect.\,2  we  describe  our
  scientific motivations.   Sect.\, 3  describes the selection  of the
  target field  and the  observations.  The data  reduction procedures
  are described in Sect.\,4 together  with the tools used to determine
  very accurate rotation periods.   In Sect.\,5 we present the results
  on the newly discovered  periodic variables.  A brief discussion and
  our plan for the near future are given in Sect.\,6.

\section{Scientific Motivations}
\subsection{Exploring magnetic activities in  T~Tauri~stars}
T Tauri stars  (TTS) are low-mass pre-main sequence  objects and based
on the strength of the H$\alpha$ line emission they are divided mainly
into two sub classes: weak  lined TTS (WTTS) and classical TTS (CTTS).
The periodic photometric variability of WTTS is linked with rotational
modulation of  cool magnetic star-spots.  Doppler Imaging,  which is a
robust stellar surface mapping technique, supports the cool spot model
(Strassmeier 2002; Schmidt et al 2005; Skelly et al.  2008).  WTTS are
found to  be relatively fast rotators with  the preliminary indication
that  they do  not  possess surface  differential  rotation (Cohen  et
al.  2004;  Skelly  et  al.    2008).   They  are  sources  of  strong
non-thermal radio as  well as X-ray emission. Since  T Tauri stars are
believed  to be fully  convective, they  cannot sustain  the so-called
$\alpha\omega$  interface  dynamo.  Furthermore,  a  fossil field  can
survive  only  over  timescales  from  10  to 100  years  in  a  fully
convective  star.  So,  even  for few-million-year-old  TTS, a  dynamo
process is necessary to generate  and amplify the magnetic field.  The
mechanism  under  which  fully  convective stars  succeed  to  produce
magnetism and related  activities is a subject of  debate (Chabrier \&
Kuker  2006  and references  therein).   The  long-term monitoring  of
low-mass  stars of the  ONC will  be very  valuable to  understand the
mechanism responsible  for the generation  of magnetic field  in fully
convective  very  young  stars.   We are  particularly  interested  to
explore  the  presence/absence  of  activity  cycles  and  of  surface
differential rotation in WTTS.

  CTTS  are surrounded  by  a circum-stellar  accretion  disk and  are
  characterised by the presence of strong emission lines as well as an
  excess  hot continuum  emission.   According magnetospheric  models,
  strong dipolar field disrupt the  inner disk at a few stellar radii.
  The disk  material is  channeled from the  inner region of  the disk
  onto  the star  along the  magnetic field  lines.  The  free falling
  material eventually  hit the stellar surface  and develops accretion
  shocks  (Calvet \&  Gullbring 1998).   The thermalized  shock energy
  form  hot spots/ring  near the  magnetic pole  which is  seen  as an
  excess blue continuum  emission in the spectrum of  cool CTTS and in
  the  optical-band photometry.  The  surface magnetic  field measured
  from Zeeman broadening in several CTTS indeed indicates the presence
  of very  strong magnetic  fields with an  average field  strength of
  $\sim$2.5~kG  (Johns-Krull  2007;  Bouvier  et al.   2007).   Recent
  results from  the Zeeman Doppler  Imaging technique also  reveal the
  presence of large-scale dipolar fields of about 1~kG together with a
  rather  complex field  configuration  close to  the stellar  surface
  (Jardine  et al.  2008;  Donati et  al.  2008).   The hot  spots are
  supposed to  be foot  prints on the  photosphere of the  large scale
  dipolar field, which facilitates  the accretion from the disk.  From
  our long-term multi-band monitoring program we intend to explore the
  temporal evolution  of the  hot spots.  In  addition, to  obtain the
  disk accretion  rate, we plan to  model the CTTS  light curves using
  star-spot models (Budding 1977; Dorren 1987; Mahdavi \& Kenyon 1998)
  combined with  the mechanism proposed by Calvet  \& Gullbring (1998)
  to account the excess emission coming from the hot spots.

\subsection{Search for PMS eclipsing binaries}
Eclipsing  binary systems  provide an  opportunity to  measure stellar
masses, radii,  effective surface  temperature, and luminosity  of the
individual  components  with  very  high  precision.   These  are  the
parameters  one  need to  test  various  theoretical PMS  evolutionary
models.   Several  monitoring programs  on  young clusters,  conducted
recently by different groups have yielded only five such PMS eclipsing
binaries (Irwin  et al.  2007;  Cargile 2008 and  references therein).
Therefore, the identification of more  such systems is indeed in great
demand.  It appears that the discovery of just five eclipsing binaries
among few  thousand of PMS stars  monitored so far,  points toward the
presence of some biases.  The first thing we must take into account is
that most of the surveys made in the recent past were single observing
run, whereas the prominent source of variability in low-mass PMS stars
are  either due to  the inhomogeneous  distribution of  cool/hot spots
and/or to the variable disk extinction. Therefore, the shallower light
variation  due to  the eclipses,  being superimposed  on  these strong
variations,  is likely  to be  masked and  only by  means  of repeated
observations  carried out  over  several seasons  one can  disentangle
these effects.  Moreover, all  PMS eclipsing systems identified so far
are  Algol-type eclipsing  variables.  The close  semidetached/contact
binaries  (if they  can form)  and partially  eclipsing  variables are
simply indiscernible  from single season  light curves, and  hence not
have been identified so far. Therefore, from our continuous multi-year
monitoring, we  expect to identify  a few more  interesting candidates
probably missed out by previous surveys.

\subsection{Studying other interesting variables}

UXor objects  are mostly intermediate-mass  PMS stars but can  also be
low-mass PMS stars.  The light curves of these stars are characterized
by sudden drops in  brightness up to 3 mag in the  V band, followed by
increased reddening and linear  polarization (Waters \& Waelkens 1998;
Grinin  et al.   1998; Herbig  2008).  In  very deep  minima  the star
reverses the color variation and,  often, becomes bluer again (Bibo \&
The  1991).  The origin  of  the  brightness  drops, the  increase  in
polarization, and the blueing effect have been debated for a long time
(Natta et al.  1997; Bertout 2000; Dullemond et al.  2003).  The model
based on variable obscuration suggest two different mechanism: passage
of proto-cometary  clouds in front of  the star (Grady  et al.  2000),
and/or  small hydrodynamic  perturbations in  the puffed-up  inner rim
(Dullemond  et al.   2003, Pontoppidan  et al.   2007).  On  the other
hand,  a  very  interesting   mechanism  was  proposed  by  Herbst  \&
Scevchenko  (1999),  in which  unsteady  accretion naturally  explains
several observable phenomena related  to UXor.  From several H$\alpha$
surveys including our own, we  find about one hundred stars surrounded
by the  active accretion disk in our  small FOV (Sect.\,5.1),  and it is
quite expected  that a few  of these will  indeed turn out to  be UXor
candidates.
  
FUor  and  EXor  are  young  low-mass stellar  objects  surrounded  by
proto-planetary accretion disk and characterized by sudden enhancement
in the luminosity.  During eruption, an FUor star can be become 100 to
1000  times more  luminous  and,  then after  it  gradually fades  up,
reaching a  quiescence phase.  On  the other hand, EXor  phenomenon is
found to be less energetic but shows more frequent/recurrent transient
activity.   Both FUor and  EXor phenomena  are generally  explained in
terms  of enhancement of  disk accretion  rate and  several mechanisms
have been  proposed to  trigger the outburst.   These include  a tidal
interaction   with  a   companion  star,   thermal   or  gravitational
instabilities,  and  induced  accretion  due to  presence  of  massive
planets (Bonnell \&  Bastien 1992; Hartmann et al.   2004; Vorobyov \&
Basu 2005;  Lodato \&  Clarke 2004).  It  is also not  well understood
whether the physical mechanism operating  in EXor outburst is the same
as  in FUors  and the  difference  in burst  duration as  well as  the
amplitude is  a consequence of  PMS evolution, or some  very different
physical process is responsible for  the EXor phenomenon. There is one
already known EXor in our field  (V1118 Ori) and we expect to identify
a few more EXor and FUor objects.

In addition  to identifying  such a new  objects, our  multi-band time
series  data  together with  the  spectroscopic  follow-up study  will
greatly  help to understand  the mechanism  responsible for  the UXor,
FUor and EXor phenomenon.

\subsection{The angular momentum evolution in the PMS phase}
In  addition to the  above mentioned  science objectives,  our program
will also be useful to address  a few important aspects of the angular
momentum evolution in  the PMS phase.  It is  believed that a fraction
of low-mass stars  surrounded by disks goes through  a phase of strong
disk-braking,  then after  the  disk-free star  freely spins-up.   The
disk-braking  models  combined  with  the process  of  star  formation
(burst/sequential)  as  well  disk  dispersal,  predict  the  bi-modal
distribution  of  the  rotation  periods  in  young  stars.   However,
numerous studies  made on  this regard over  the last  decade ended-up
with  very  contradictory  claims   and  counter  claims  (Stassun  et
al. 1999; Herbst et al.  2002;  Rebull 2004; Lamm et al.  2005, Rebull
2006;  Cieza \&  Baliber 2007).   In most  studies of  stellar angular
momentum, the  stellar rotation period  is obtained from  light curves
produced  by  the rotational  modulation  of  the  star light  due  to
inhomogeneous  cool/hot  spots, unevenly  distributed  on the  stellar
surface.   However, there  have been  no  attempts made  to check  the
completeness of the rotation periods  derived from the light curves of
various objects (except some work done  by Cohen et al. 2004 on IC348;
Lamm et al. 2005 on NGC2264).  It is well known that the cool spots on
WTTS and other  active stars can change their size  as well as spatial
distribution so  dramatically and rapidly  that one can not  expect to
get always the rotation modulation, specifically when the amplitude of
variation  is  small.   That  is  the reason,  for  example,  why  the
single-season survey made  by Herbst et al.  (2002)  in 1998-1999 with
the MPG/ESO  2.2m telescope could  not detect all the  bright periodic
variables identified by the previous survey of Stassun et al.  (1999),
despite using  a larger telescope and, hence,  with better sensitivity
and accuracy.  On  the other hand, stars with  disks (mainly CTTS) are
supposed to have both cool and hot spots.  These stars usually display
very irregular  light variations  and, often, it  is difficult  to get
their rotation period accurately, at least from a single observing run
(Herbst et al. 1994; Grankin et al. 2007). One of the question we want
to address is why just  about 10-20\% of several thousands of low-mass
PMS stars monitored  so far are found to  be periodic variables.  What
happened to the remaining 80-90\% of  stars?  Why do they not show any
regular and  periodic variation despite having all  the ingredients to
produce  magnetic   spots?   So,  our   long-term  monitoring  program
hopefully  will  put  us  in  the  position  to  shed  some  light  on
detectability  of  the rotation  periods  and  the  effects of  biases
preventing their detections.

\section{Selection of the field and Observation}

\subsection{The selection of the  field}
\begin{figure}
\begin{center}
\epsfig{file=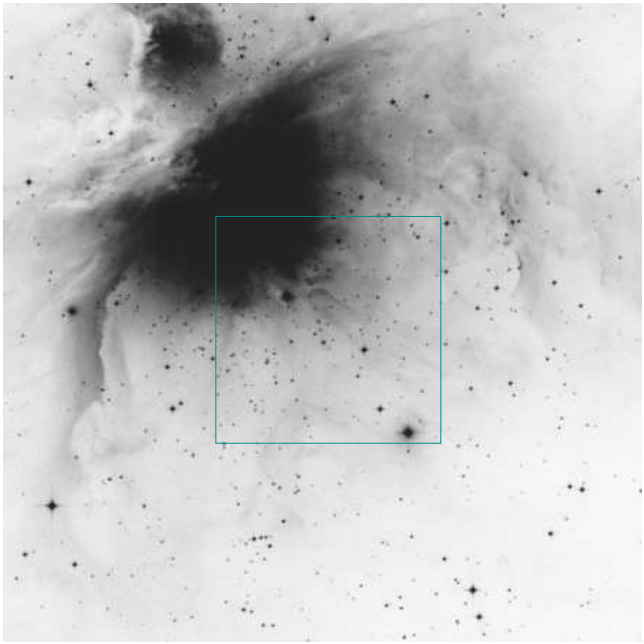,scale=0.6}
\caption{A $30\times30$ arc-min red band  image from DSS on which   the 
   $10\times10$ arc-min region chosen for the long-term monitoring  is marked.}
\label{onc_dss}
\end{center}
\end{figure}

 Keeping in mind the science goals mentioned in the preceding section,
 we started  looking for  a stellar field  associated with  very young
 stellar  clusters   containing  a  large  number   of  confirmed  and
 relatively bright  young members confined to a  small region. Another
 criterion  was that the  field must  have a  large number  of already
 known PMS variables, representing to different mechanisms responsible
 for the light variations.  Our search finally ended on the ONC, which
 was  already chosen  as target  by other  major  monitoring programs,
 conducted by  Stassun et  al.  (1999), Herbst  et al.  (2000), Rebull
 (2001), Carpenter  et al.  (2001),  and Herbst et al.   (2002).  More
 importantly, since  1990 hundreds  of bright stars  of ONC  have also
 been  continuously  monitored  by  Herbst and  his  collaborators  of
 VVO. The field of view (FOV) of the telescopes what we planned to use
 was  10$\times$10  arc-minutes.  Since  our  program needs  long-term
 monitoring at  least in two  broad-band filters, and  considering the
 limited  availability  of  the  telescope  time, we  had  to  further
 optimize a very potential 10$\times$10 arc-minutes field having large
 number of variables along with other measured quantities, such as NIR
 as well  as X-ray  data.  We  identified a region  south west  of the
 Trapezium stars shown in  Fig.\,\ref{onc_dss}.  The coordinate of the
 selected    ONC   field    is    $\alpha$(2000.0)   =    05:35:04.09,
 $\delta$(2000.0)  =  $-$05:29:04.8.  We  deliberately  kept the  very
 bright trapezium stars out of the field, so that the strong scattered
 light and very severe bleeding of the charge should not spoil the CCD
 frame. The  ONC field  chosen by us  includes 110  periodic variables
 already  known from  the literature:  92 from  Herbst et  al.  (2000,
 2002) and 38 stars from Stassun et al.  (1999, hereafter called S99),
 where  20  periodic  variables   are  common  in  both  groups.   The
 cross-correlation  of  optically-identified  stars  from  Hillenbrand
 (1997), with 2MASS  all sky catalogue of point  sources (Cutri et al.
 2003) and COUP X-ray observation (Getman et al.  2005), revealed that
 almost all  optical stars have  NIR counterparts and large  number of
 them have  X-ray data  as well.  Most  of these objects  are low-mass
 stars and extremely young candidates. Almost one fourth of our FOV is
 so close to Trapezium and severely  affected by a strong HII region (
 see  Fig.\ref{onc_dss}), whose  average background  level  is several
 times larger than in the  region slightly away from it.  Nonetheless,
 we included this region because it hosts a large number of relatively
 young low-mass stars (younger than  1~Myr) and most of them have been
 found to  be source of  strong X-ray emission.  Furthermore,  it also
 comprises  a  large number  of  intermediate-mass stars  (Hillenbrand
 1997).

\subsection{Photometric Observations}
\begin{figure}
\begin{center}
\includegraphics[scale = 0.45, trim = 40 00 00 00, clip]{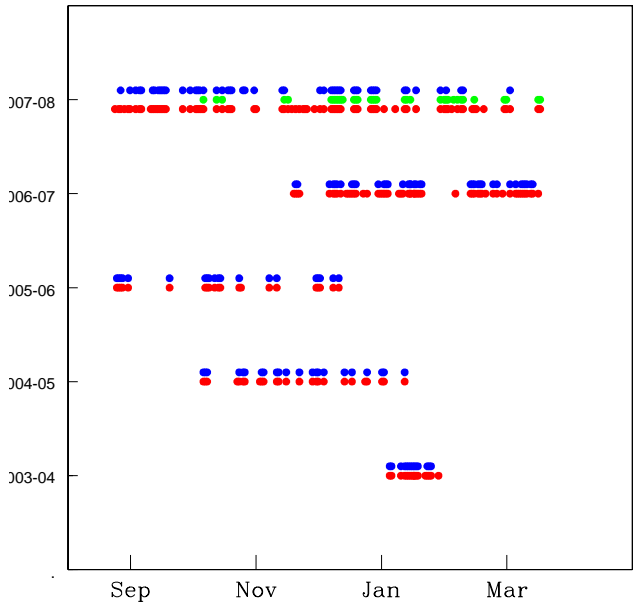}
\caption{The distribution of the observed nights for all five cycles.   
The red,   green and blue   colors represent   the observation made with I, R and V
   filters, respectively. }
\label{obsnight}
\end{center}
\end{figure}

 The photometric observations of  the ONC reported here, were obtained
 between January 18,  2004 and April 15, 2008,  using the 2m Himalayan
 Chandra  Telescope (HCT) and  the 2.3m  Vainu Bappu  Telescope (VBT).
 The Himalaya Faint Object Spectrograph (HFOSC) of HCT used in imaging
 mode uses 2k$\times$2k central  region of 2k$\times$4k CCD and covers
 a field  of view of $10\times10$  arc-minutes, with a  scale of 0.296
 arc-sec/pixel. The  $1k\times1k$ CCD mounted  at the VBT  prime focus
 covers a slightly larger field ($\sim$ 11$\times$11 arc-minutes) with
 plate scale of 0.65 arc-sec/pixel.  The time series observations were
 made  primarily through Bessel  I \&  V broad-band  filters. However,
 during  the most  recent observing  run (2007-08)  we  collected time
 series observations  in R  band too.  The  most complete  time series
 data  in all  observing  runs is  in  I band.  Reasons  to give  more
 emphasis  to I  band are:  (i)  the I  band minimizes  the effect  of
 nebular background and interstellar  extinction, as well it maximizes
 the S/N of  red faint stars which constitute the  bulk of the cluster
 population, (ii) past monitoring programs were conducted primarily in
 I band  and so  the comparison and  the use  of earlier data  is only
 possible with I  band, and finally (iii) the effect  of the seeing is
 least  in I  band,  and  the effect  of  color dependent  atmospheric
 extinction,  which is difficult  to correct,  is also  minimized.  In
 order to avoid the degradation of seeing at low elevation, effort was
 also  made  to  obtain  photometric  observations  at  low  air-mass.
 Whenever  possible, immediately  after I  band observations,  we also
 tried to  collect V \& R band  data with motivation to  get the color
 curves of  at least a few  bright and less  embedded cluster members.
 During  the  observing  rus  2004-06,  whenever  telescope  time  was
 available, a sequence of 3-5 frames  in each I and V filters with the
 exposure  time of 60  and 180  seconds were  collected.  In  order to
 increase  the dynamic range  as well  as the  photometric measurement
 accuracy of faint stars, starting  from the most recent observing run
 (2007-08) we changed the observing strategy.  Now, one short exposure
 is accompanied by 3-4 long exposures (20 and 90 seconds in I band, 60
 and 300 seconds in V and R bands).  Whenever possible, this observing
 sequence  which finally gives  one averaged  data point  was repeated
 more than once per night.  In total the ONC field was observed on 208
 nights and collected 1986 frames in R, V and I bands.  A brief log of
 our observations is given  in Table~\ref{obslog} and the distribution
 of  the  nightly  observations  for  all  the  filters  is  shown  in
 Fig.~\ref{obsnight}.   The   median  seeing  in  I   band  for  these
 observations is  \near 1.5 arc-sec  (FWHM) and, during the  course of
 our  observations, it  varied from  0.8  to 2.5  arc-sec.  Seeing  at
 shorter wavelengths was slightly degraded according to the law of the
 wavelength dependent seeing (FHWM$\propto \lambda^{-1/5}$).

  Since  accurate   flat  fielding   is  of  critical   importance  in
  differential photometry  too, therefore, on every night  we tried to
  collect a  large number of evening  and morning sufficiently-exposed
  twilight flats.  Since there is  no over-scan region in the detector
  used  by  us, several  bias  frames,  spread  over the  night,  were
  collected.  It  has been found that  even after taking  all sorts of
  precaution to  do very accurate  flat fielding, small errors  of the
  order of  0.1 to 1.0\%  remain in the  flat field (primarily  due to
  non-uniform illumination of  the flat field and wavelength-dependent
  differential  variation in  the quantum  efficiency of  the pixels).
  Furthermore, our back-illuminated thin  CCD chips on both telescopes
  suffer from high spatial frequency fringing problem. The combination
  of  these two  effects typically  limits the  achievable photometric
  accuracy to a few milli-magnitude (mmag) depending on the instrument
  used. In  order to minimize these effects,  throughout the observing
  run, we tried to keep the stars  in our FOV at the same pixel on CCD
  chip.   To  do  this we  selected  a  moderately  bright star  as  a
  reference star and kept this star at reference pixel just before the
  closed loop guiding used to  begin. During 90\% of observations, all
  target objects were kept within the 2-3 pixels on the CCD frames.

\begin{table}
\caption{\label{obslog} Log of the observations made to date.}
\begin{center}
\begin{tabular}{|c|c|c|c|c|}
\hline
 Cyc Start    & Cyc End     & Filter  & No. Frames &  No. Nights \\
\hline                        
Jan 18, 2004  & Feb 16, 2004& I  &   121   & 16 \\
              &             & V  &    69   & 14 \\
Oct 11, 2004  & Jan 26, 2005& I  &   109   & 25 \\
              &             & V  &    80   & 22 \\
Aug 26, 2005  & Dec 22, 2005& I  &    89   & 21 \\
              &             & V  &    58   & 19 \\
Nov 28, 2006  & Apr 07, 2007& I  &   337   & 57 \\
              &             & V  &   159   & 38 \\
Aug 25, 2007  & Apr 18, 2008& I  &   492   & 89 \\
              &             & V  &   274   & 35 \\
              &             & R  &   198   & 50 \\

\hline
\end{tabular}

\end{center}
\end{table}

\subsection{Slit-less spectroscopy}
In order to identify stars which  show H$\alpha$ in emission, in a few
observing nights  of 2006-07, when  the seeing was  relatively better,
the ONC field was also  observed using HFOSC in the slit-less spectral
mode with a  grism as dispersing element.  In  this mode a combination
of the  H$\alpha$ broad-band  filter (H$\alpha$-Br, 6100  - 6740$\AA$)
and Grism~5  were used without any  slit.  This yields  an image where
the stars are  replaced by their low-resolution spectra,  which are on
average  displaced  by  163  pixels  upward in  the  CCD  plane.   The
2048$\times$2500 pixels,  central region  of the 2k$\times$4k  CCD was
used for  data acquisition.  The average dispersion  of Grism~5 around
H$\alpha$  is 3.12~\AA/pixel  and the  seeing was  around  1.2 arc-sec
($\sim$4  pixels), which  sets  the system  resolution  at about  500.
Several repeated observations with 600  second were made on each night
and bias as  well as twilight flats were also taken  on the same night
with the same setup.  After  bias correction and the flat fielding, as
done in the  optical imaging, the spectral images  were combined using
median  combine.  The  consecutive  spectral images  were taken  while
auto-guiding   was  active,  so   image  shifts   were  found   to  be
insignificant and  stellar spectra after median combine  were found to
be not suffering  through any smearing effect.  On  the other hand, we
could improve not  only the signal to noise ratio  (S/N) in the median
combined spectral  image, but  also remove the  cosmic ray  events and
hence  ensure unambiguous determination  of sharp  H$\alpha$ emission.
From the  median combined  image, strips of  the image  containing the
stellar spectra were copied and then the optimal extraction procedures
carried out in slit-spectroscopy were followed.  We used the IRAF task
\textit{apall} to extract  the spectra of 346 stars.   A very accurate
sky subtraction from those stellar spectra which have been affected by
the strong HII region is indeed very crucial.  We sampled the sky from
2.5 arc-sec wide sky regions,  nearly 2.5 arc-sec away, on either side
of  the stellar spectrum.   To over  come the  problem of  small scale
strong  variation in the  sky background,  a low-order  polynomial was
used  to fit  the  background across  the  dispersion. This  technique
ensures  that the  nebular contribution  in the  H$\alpha$  profile is
negligible and  the measured H$\alpha$ emission is  indeed coming from
the star.   The coarse  wavelength calibration was  done by  using the
average dispersion of  the Grism~5, where the center  of the H$\alpha$
line  was  used  as  a  reference point  (6562.8\AA).   The  H$\alpha$
emission line stars were identified and the equivalent widths(EW) were
measured.  The  smallest EW  measured from our  slit-less spectroscopy
was  found to  depend on  both  seeing condition  and star's  apparent
magnitude.   The  later fixes  the  strength  of  the continuum,  with
respect to  what we measure the  EW of emission lines.   On the seeing
value  smaller than  1.5  arc-sec, we  could  measure EW  as small  as
1\AA\,\,\ and this can also be  considered the typical error in our EW
for the stars having well exposed continuum.

\section{Data reduction}
\subsection{Pre-processing and selection of the target stars}
The   basic   image  processing,   such   as   bias  subtraction   and
flat-fielding,  were done  in  a standard  way  using tasks  available
within  IRAF. In  order  to minimize  the  propagation of  statistical
errors  while doing  the bias  subtraction  and flat  fielding, it  is
recommended that on  each night one need to collect  a large number of
bias and  flat frames  . However,  on average not  more than  4-8 bias
frames  as well  as twilight  frames could  be collected,  due  to the
constraint imposed  by long read-out  time ($\sim$ 80~sec) of  the CCD
used.   Therefore,  to  minimize  the  effect of  propagation  of  the
statistical  error  a  large  number  of bias  and  flat  frames  were
collected from near-by  nights and, depending on the  stability of the
features on  the frames, bias and  flats of 3 to  4 consecutive nights
were combined to construct the  master bias and flat frames. After the
bias correction and the flat fielding, the next task was to identify a
large number of sufficiently bright stars  in our field. To do this we
selected one  I-band frame  which was obtained  at fairly  good seeing
condition and used the \textit{daofind} task of IRAF to identify stars
within it.   The radial profile  of individual object was  checked and
the  false  detections  or  profile featuring  extended  objects  were
rejected.  A total of 346  objects suitable for time series study were
identified  in this  reference  frame.  Although  the  whole frame  is
affected to some extent by nebulosity, however, based on the intensity
of the  background emission  we divided our  FoV into  three different
sections, clear sky,  sky partially affected by nebula  and the region
where the nebula  is very strong.  The stars  falling in these regions
were marked with the sky flags, SC, SPN and SN.  In addition to this a
SNB flag  is used  for bright  stars located in  the region  where the
nebular emission is strong.  Nearly 5\% of the frames taken under very
poor sky transparency or worst  seeing condition or affected by strong
wind were rejected for subsequent analysis.

The astrometric  calibration of the stars identified  in the reference
frame  discussed above  was done  using the  \textit{Starlink} package
ASTROM (Wallace 1994).  The calibration was done using the coordinates
of about one hundred stars listed in the 2MASS catalogue (Cutri et al.
2003) as  references and on  average the accuracy achieved  on stellar
coordinates is  about 0.2 arc-sec.   From the comparison  with various
coordinates  reported  by earlier  researchers,  it  appears that  our
coordinates  very  well  match  with  the  coordinates  of  Herbst  et
al.  (2002),   with  average  difference   of  0.23$\pm$0.21  arc-sec.
Whereas,  systematic 0.68$\pm$0.15  arc-sec and  1.39$\pm$0.28 arc-sec
differences have been noticed with Hillenbrand (1997) and S99.  Herbst
et al.  (2002) also reported  such difference and  it was found  to be
systematic  errors in  the astrometric  transformation carried  out by
above two surveys.

\subsection{The photometry}

Despite our  effort to keep our  program stars always at  the same CCD
pixel, the observed frames were found to be typically out of the place
and  shifted  with   respect  to  the  reference  frame   by  1  to  3
pixels. Furthermore, there were about  10\% of the frames in which the
effort was  not made to center  the star at the  reference position on
the CCD.   So, before  we start the  photometric reduction we  had two
options. One possibility was to align all the frames with respect to a
reference  frame  and  to  use  the  same  coordinates  obtained  from
reference  frame to  all the  frames.   The other  possibility was  to
convert the  coordinates of the stars determined  from reference frame
taking the image shift as well as the rotation into account.  We opted
the  later procedure because  the first  procedure uses  the intensity
interpolation for the fractional shift  in pixels and may not conserve
the flux. The  center of tens of bright stars  were obtained using the
CCD  data reduction  package  DAOPHOT-II (Stetson  1987; 1992).   Then
after,  the  Stetson's  \textit{daomatch} and  the  \textit{daomaster}
programs were  used to obtain reasonably  good transformation relation
of  stellar  positions  between  the  reference frame  and  any  other
observed frame.   In the  subsequent step the  transformation relation
generated  by  \textit{daomaster}  was   used  to  generate  an  input
coordinate file of  all 346 stars whose coordinates  have been already
determined in the  reference frame.  This coordinate file  was used as
input to  either aperture photometry carried out  by the \textit{phot}
task of IRAF or the PSF photometry done using Stetson DAOPHOT-II.

For each  star we performed both  aperture as well  as PSF photometry.
In aperture  photometry, the magnitude  of stars were  determined with
aperture radius spanning the range of 4-10 pixels.  Keeping relatively
poor seeing in consideration (FWHM varies from 4 to 9 pixels), the sky
was estimated from an annulus  with slightly larger inner radius of 30
pixels ($\sim8.8$  arc-sec) and  width of 10  pixels.  We  played with
various sky estimation procedures available in the IRAF and found that
the  \textit{mode}  gives more  precise  sky  value  and, hence,  less
scattered light curves of  non-variable stars.  Therefore, the sky was
always estimated by using the \textit{mode} option, whenever, aperture
photometry  was carried  out.   The PSF  photometry  was performed  by
modeling the  star's profile  with a Penny-2  function.  We  used this
function because we found it  yield a least residual after subtracting
fitted stars from the  image.  Besides the analytical functions, which
are used for the fitting procedure, another important parameter of PSF
photometry  is   the  fitting-radius.   For  each  frame   we  used  a
fitting-radius equal to the mean FWHM of the stellar profiles.  Such a
value was  about 4 pixels for  the images acquired in  good seeing and
about 8  pixels in  poor seeing conditions.   If any faint  object has
bright neighbours, then the effects of variable seeing combined to the
intrinsic variability of the  bright neighbours can introduce spurious
variations.  Such  close pairs were  identified and their  time series
photometric  data  was  treated   with  special  care.   On  total  we
identified 17 such pairs which have been found to be separated by less
than 6 arc-sec.

\subsection{Ensemble differential photometry}

The  differential  photometry was  performed  using  the technique  of
ensemble photometry (Gilliland \&  Brown 1988; Everett \& Howell 2001;
Bailer-Jones \& Mundt 2001).   In ensemble photometry the differential
magnitude is computed with respect to the average magnitude of a large
number  of  non-variable  reference  stars.   The  advantage  of  this
technique is that, in the  averaging process, the uncertainties of the
ensemble stars  magnitudes due to statistical fluctuations  as well as
short-term small  incoherent variations  will cancel each  other.  The
uncertainty in  the magnitude of the  artificial comparison, therefore
will be smaller than the uncertainty on the magnitude of a single star
and this, in turn, will produce less noisy light curves.  The ensemble
photometry was performed with ARCO, a software for Automatic Reduction
of CCD Observations,  developed by us (Distefano et  al.  2007).  This
software allows us to select  automatically the suitable stars for the
ensemble,  i.e., sufficiently  bright  and isolated  stars, which  are
common to all  the frames, and distributed all over  the frame but not
found  to be close  to the  CCD edges.   After selecting  the ensemble
stars,  ARCO   automatically  computes:  (i)  the   magnitude  of  the
artificial comparison star by  averaging instrumental flux of ensemble
stars, (ii)  time-series differential magnitudes for each  star of the
field  and (iii) mean,  median and  the standard  deviation ($\sigma$)
associated to each time  series.  While computing the average ensemble
magnitude, we first determined the  average flux of all ensemble stars
and, then after, the ensemble  magnitude was computed from the average
flux. This  way of computing  ensemble magnitude gives more  weight to
the bright stars  which are expected to have  smaller error related to
photon noise.

The  whole procedure  described above  is iterative  and start  with a
large number of bright stars distributed all over the frame, excluding
stars very  much affected by  the nebulosity.  While  constructing the
ensemble,  the program  exclude those  ensemble stars  whose standard
deviation is larger than the  threshold sigma value fixed by the user.
The  final  output  of  the  software  is  a  file  with  differential
time-series  magnitudes  for all  stars  in  the  field including  the
ensemble  stars.  We  used the  ensemble made  up of  24 stars  to get
differential magnitudes in  I, R and V photometric  bands.  We run the
program several times with  different input data coming from apertures
as  well from  PSF  photometry  and generated  several  sets of  light
curves.

In principle, when differential  aperture photometry is performed, the
choice of the radius should not matter because, if the stellar profile
does not  vary significantly across  the frame, the percentage  of the
total flux collected through an aperture of a given radius $r$ is same
for all  stars and, therefore,  there should be no  difference between
time-series data  obtained with a  4-pixel aperture radius or  with an
8-pixel radius.  However,  the use of a fixed radius  for all stars is
not recommended  when the  stars of  the field span  a broad  range of
magnitudes.  Although a larger aperture increases the signal strength,
however, at the same time,  it increases the contribution of the noise
due to the  background fluctuations and CCD read out.   In the case of
bright stars the contribution to  the total noise is mostly the photon
noise whereas, the noise in the faint star magnitude is due to the sky
and read out noise of the  CCD.  So, a large aperture is preferable to
bright stars and a smaller aperture  to faint stars.  It is well known
that signal-to-noise ratio is a  function of size of the apertures and
is found  to be maximum close the  aperture radius of one  FWHM of the
stellar profile  (Howell 1989).  The  magnitude of faint stars  may be
very much affected even by  slight inaccurate estimation of sky values
and this can  be minimized by adopting a  small aperture. Whereas, any
variation in the  shape of PSF over the frame  introduces error in the
bright  stars  magnitudes. Keeping  all  these  in  mind, a  range  of
apertures  starting from  4  to 8  pixels  were used  to generate  the
ensemble differential photometry.

\begin{figure}
\begin{center}
\epsfig{file=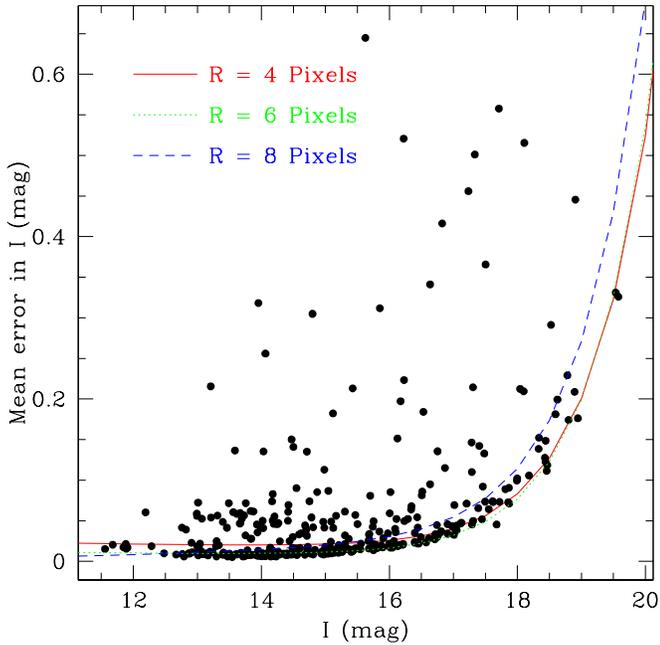,scale=0.45,angle=0}
\caption{The mean error associated to each data points as function 
of the I band magnitude. The sequence which comprises the  maximum 
distribution of the data points, were fitted with a piecewise function 
of second order polynomials and the exponential function. The data points 
of  aperture radius 6 has been only plotted here, but fitting were done 
for data of all three apertures  4, 6 and 8 pixels radius and the best
 fit curves are shown here. Most of the deviant data points with respect to the fit, are associated with the stars in strong nebula and hence the background photon noise are 
the prime source of the error.}
\label{merror}
\end{center}
\end{figure}

As mentioned in Sect.\,3,  the differential ensemble magnitudes of the
sequence of  3 to  4 frames collected  within very short  intervals of
time (shorter  than 1 hour)  were combined and  the mean value  of the
magnitudes from  these close-by frames was  used as one  data point in
the time series analysis. We also computed standard deviation of these
magnitudes which  is a  robust estimate of  the error  associated with
each data point.  The mean of the standard  deviation $<\sigma>$ is an
average error of  the measurement associated with any  star which is a
function of the magnitude  and plotted in the Fig.~\ref{merror}.  Such
an  estimate of  error  linked  with one  data  point is  conservative
because the  true observational accuracy could be,  in principle, even
better for  stars having substantial variability  within the timescale
close  to our fixed  binning time  interval (i.e.   1 hour).   We have
computed the mean  error $<\sigma>$ for all three  apertures of radius
4, 6 and  8 pixels, respectively. The lower bounds  of the data points
were fitted with a piecewise  function of second order polynomials and
the  exponential function (see  Fig.~\ref{merror}).  As  expected, the
smaller  apertures  give better  photometric  precision  to the  faint
stars.  Whereas, the large apertures  seem to be more suitable for the
brights stars.  Relatively large errors associated with the magnitudes
determined using small aperture of  bright stars reflect the effect of
temporal as  well as  spatial variation of  the PSF.   The photometric
precision achieved in the interval of 11.5$<$I$<$16.0 is close to 0.01
($\sim$1\%) and then it  degrades exponentially.  During the final run
of  the  photometric reduction,  the  optimum  aperture  was not  only
selected based on  the stars brightness, but the  aperture which gives
the lowest mean error ( $<\sigma>$) was also taken in to account.

\begin{figure}
\begin{center}
\epsfig{file=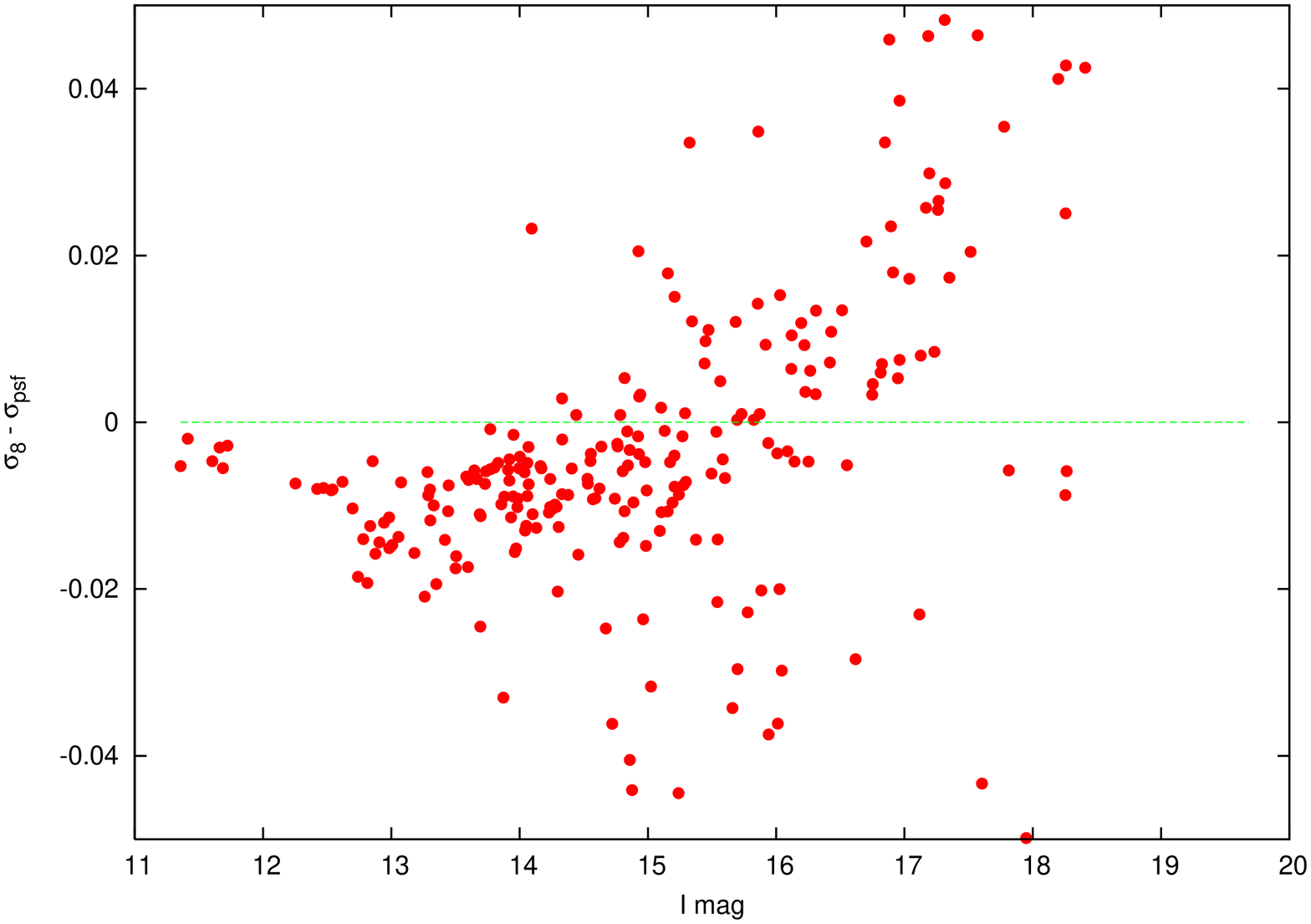,scale=0.3,angle=-90}
\epsfig{file=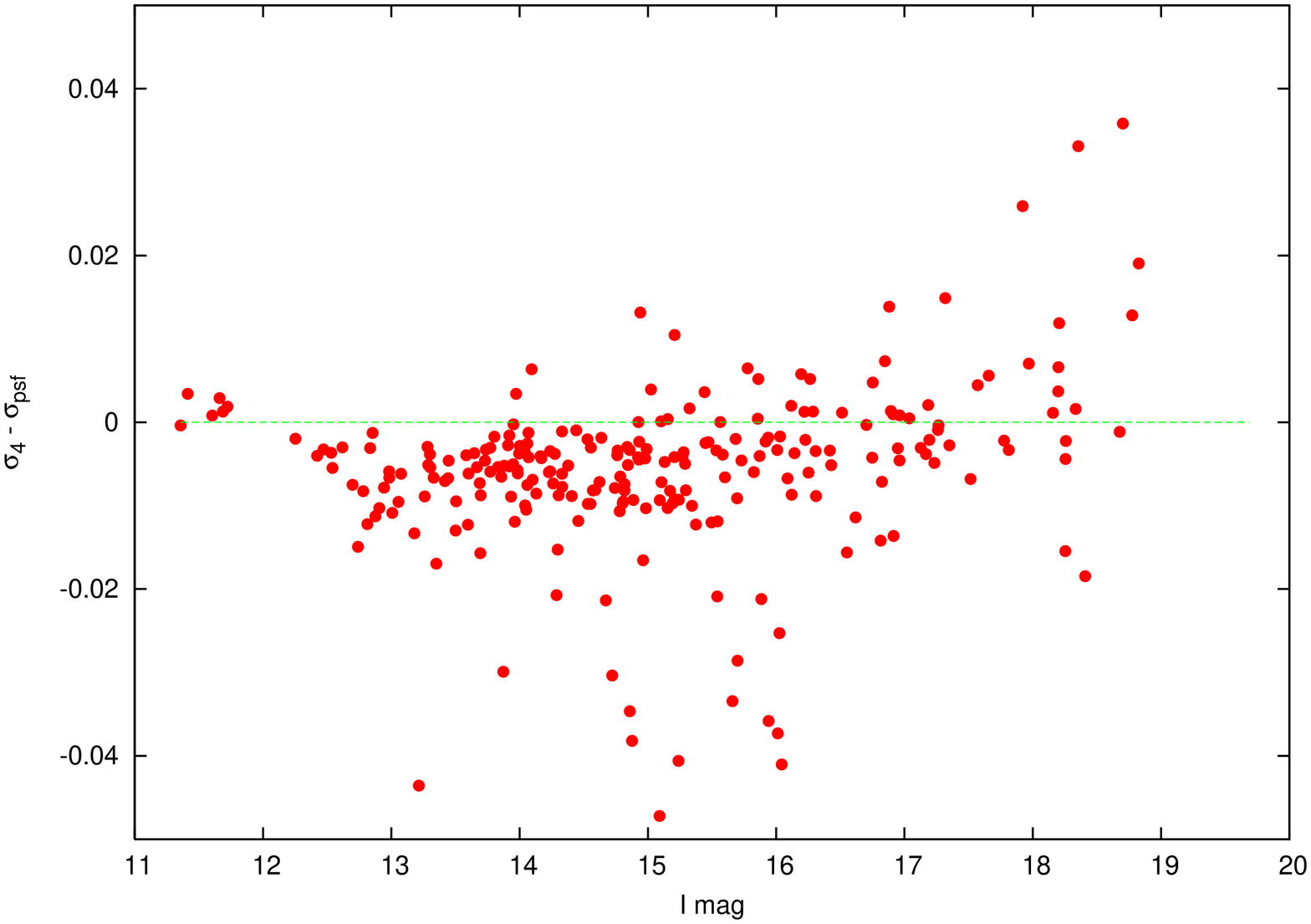,scale=0.3,angle=-90}
\caption{In the top panel the quantity "$\sigma_8-\sigma_{psf}$" vs. the I magnitude is plotted. The radius of 8 pixels gives a smaller standard deviation and, in turn, a less noisy light curve for  brighter stars. To the fainter stars PSF photometry seems to be advantageous, but looking at the $\sigma_4-\sigma_{PSF}$ vs. I~mag plot (bottom panel), it appears  that also in such a case aperture photometry gives the better results.}
\label{conf.ap8.ps}
\end{center}
\end{figure}

Finally, we  performed a comparison between the  results obtained from
aperture  photometry and  from PSF-photometry.   Such a  comparison is
shown  in  Fig.~\ref{conf.ap8.ps},  where $\sigma_8-\sigma_{psf}$  and
$\sigma_4-\sigma_{psf}$~vs.~I~mag are plotted.   In the fainter domain
PSF-photometry is  more advantageous than  aperture photometry carried
out  with 8-pixel  radius (Fig.~\ref{conf.ap8.ps}).   However,  if the
aperture photometry is done with  radius of four pixels, then aperture
photometry    gives   better   results    in   the    fainter   domain
(Fig.~\ref{conf.ap8.ps}).  From all these  detailed exercise  we found
that  in   our  case  generally   the  aperture  photometry   is  more
advantageous than  PSF-photometry.  Nevertheless, we  found that there
are  few stars  for  which PSF-photometry  produces  less noisy  light
curves and we  noticed that such stars are either  close to a brighter
star or lying close to edges of the CCD.  In such cases PSF-photometry
is  more efficient  because  it allows  us  to take  into account  the
effects of the  distortion of the stellar profile at  the edges of CCD
as  well  as it  "deblends"  the  star  from the  brighter  neighbours.
Therefore,  to construct  time  series  data of  these  stars we  used
PSF-photometry.

 \subsection{Absolute Photometry}

In  order to  obtain the  standard magnitudes  and the  colors  of all
targets  in our  FOV, which  allow  our observations  to compare  with
previous observations, we decided to carry out photometric calibration
as precise as possible. The standardization of magnitudes also enables
us to correctly  place our objects in the HR  diagram and to determine
various stellar  parameters.  On  the six best  nights of  the 2007-08
observing run, we observed a large number of BVRI photometric standard
stars from Landolt  (1992) and deeper asterism of  M67 (Anupama et al.
1994).   A few  Landolt fields  were monitored  over a  wide  range of
air-mass to  determine the  nightly atmospheric extinction.   From the
extinction observation we found that  only four out of six nights were
photometric  and the  transformation coefficients  were  obtained from
these nights.   On the same  night a large  number of VRI  frames with
short and long  exposures of the ONC were taken, when  it was close to
the meridian.   The short  and long ONC  frames were aligned  and then
co-added separately, using the median  option of the \it imcombine \rm
task  of IRAF.   The aperture  photometry magnitude  with a  radius of
1.5$\times$FWHM, which is suppose to  give maximum S/N was carried out
for  the all  346 ONC  variables.  Then  after, 20-30  fairly isolated
bright stars free from nebulosity  were used to determine the aperture
correction.   Finally the  aperture  corrected magnitudes  of the  ONC
stars were obtained for the  aperture of 5$\times$FWHM and then I-band
magnitude  and  colors   were  transformed  using  the  transformation
equation

\begin{equation}
         I   =  i - k_i X  + \epsilon  (R-I) + \zeta_i 
\end{equation}
\begin{equation}
         (R-I)  = ((r-i)-k_{ri}\Delta~X)   \mu_{ri} + \zeta_{ri} 
\end{equation}
\begin{equation}
         (V-I)  = ((v-i)-k_{vi}\Delta~X)   \mu_{vi} + \zeta_{vi} 
\end{equation}

\noindent where  k, X, $\epsilon$,  $\mu$ and $\zeta$  are atmospheric
extinction,  air-mass, transformation  coefficients, and  zero points,
respectively.  The average magnitude and  colors of all 346 stars were
obtained from four nights using Eq. (1-3).

\begin{figure}
\begin{center}
\includegraphics[scale = 0.80, trim = 30 160 160 00, clip]{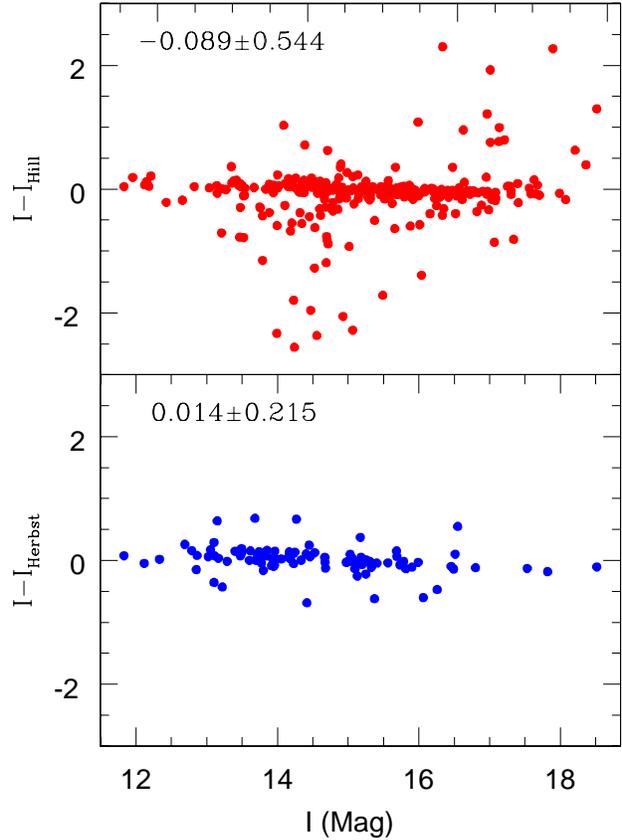}
\caption{ The comparison of   I-band   magnitudes obtained from our photometry and   from  Hillenbrand (1997), and  Herbst et al. (2002).   The mean  and the standard   deviations   of the difference in magnitudes  are also given on the top of each plots.}
\label{onc_imag_comp}
\end{center}
\end{figure}

Finally, the accuracy of  the photometric calibration was estimated by
computing the  standard deviation of  magnitudes and colors of  the 24
comparisons used  for the  differential ensemble photometry.   And the
errors were  found to be 0.02, 0.04,  and 0.05 mag in  I band, (R$-$I)
and (V$-$I)  colors, respectively.  Because,  nearly all stars  in our
field are expected to be variable with amplitude of variability in the
range from  our accuracy  limit (0.01mag) to  few tens  of magnitudes,
therefore,  a  better estimate  of  their  brightness  comes from  the
mean/median value  of the time series data.   Therefore, we determined
the median magnitudes  of each star's time series  data collected over
five consecutive  observing years and  compared these values  with the
I-band magnitudes of Hillenbrand (1997)  and Herbst et. al (2002), the
latter collected during a complete observation season.  The difference
of the  median magnitudes obtained  from the common stars  are plotted
against our  I-band magnitude in  Fig.~\ref{onc_imag_comp}. Our median
magnitudes and those from Herbst  et al. (2002) seem to matching well,
whereas, the  difference with respect  to Hillenbrand (1997)  is quite
apparent in the plot.  Here we remind that, differently than Herbst et
al.   (2002), the  photometric  data used  by  Hillenbrand (1997)  was
mostly based  on snapshot observations  collected over few  nights and
hence  affected by  the intrinsic  variability.  The  photometric data
along with other relevant information of  all 346 stars in our FOV are
partly given in Table~\ref{basic_data}, whereas, the complete table is
available only electronically.

\begin{table*}

\centering
\begin{minipage}{\linewidth}
\caption{The photometric data along with other relevant information of all 346 stars in our FoV.\label{basic_data}}
\scriptsize
\begin{tabular}{lrrrrrrrrrrrrrrr}


\hline\hline
S.N. & RA-Dec           &  JW &  P$_{Herbst}$ & P$_{Stassun}$ & Sky  & Neighbour & CTTS & I & R-I & V-I & Sp.Type & J & J-H & H-K & L$_X$\\
\hline 
 161 & 05 34 55.006 -05 26 58.90    &  3111   &    -     &    -         &	SC	&-     &  - &   16.07     &     1.44    &     3.22 &   M5.5e     &  13.83&    0.92&    0.61&  29.00  \\
 162 & 05 34 53.100 -05 26 59.54    &   -     &    -     &    -         &	SC	&-     &  - &   18.61     &     3.60    &     2.67 &   -         &  15.96&    0.79&    0.35&  28.04  \\
 163 & 05 34 50.727 -05 27 01.01    &  117    &   8.870  &    -         &	SC	&-     &  C &   13.17     &     1.02    &     2.10 &   M0e       &  11.65&    0.93&    0.53&  29.26  \\
 164 & 05 35 21.627 -05 26 57.78    &  688    &    -     &    -         &	SPN	&-     &  - &   14.68     &     0.74    &     2.81 &   -         &  12.84&    0.76&    0.31&  29.10  \\
 165 & 05 35 05.851 -05 27 01.64    &  281    &   3.160  &    -         &	SPN	&-     &  - &   13.95     &     1.78    &     3.29 &   M3.5      &  12.10&    0.96&    0.32&  29.56  \\
 166 & 05 35 06.426 -05 27 04.82    &  290    &    -     &    -         &	SPN	&-     &  - &   15.37     &     1.56    &     3.81 &   -         &  12.93&    0.74&    0.48&  29.41  \\
 167 & 05 35 05.891 -05 27 09.00    &  283    &   7.010  &    -         &	SPN	&-     &  - &   15.19     &     0.48    &     2.79 &   K8        &  12.51&    1.37&    0.75&  29.72  \\
  \hline
\multicolumn{16}{l}{NOTE:  Only  a portion of the table is shown here and complete table is  available only in  electronic edition of the MNRAS.}\\

 \end{tabular}
\end{minipage}
\end{table*}

\subsection{Search for periodicity}

As already mentioned, one major  objective of our project is to detect
and characterize the  optical and NIR band variability  of all targets
detected  in  our  FOV.   Specifically,  we  aim  at  discovering  new
variables   and  their   rotation  period   whenever   possible.   The
variability  of low-mass  members  of ONC  mostly  arises from  uneven
distribution  of cool/hot  brightness inhomogeneities  on  the stellar
photosphere, which,  being carried  in and out  of view by  the star's
rotation,  produce a  quasi-periodic variation  in the  observed flux.
The variation in the star's  light is modeled through Fourier analysis
to  determine  the  stellar  rotation  period.   There  are  transient
phenomena  related   to  magnetic   activity,  such  as   flaring  and
micro-flaring, and star-disk interaction  which also give rise to flux
variability.  However, they generally  tend to be non-periodic, making
more difficult the  detection of any periodicity in  the observed time
series  data.   The reliable  determination  of  the stellar  rotation
period  is possible  if several  conditions are  met at  a  time.  For
example, stars must  have surface inhomogeneities unevenly distributed
along  the  stellar  longitude,  the  stellar  latitude  of  spots  in
combination with the  inclination of the rotation axis  must allow the
rotation  to  modulate  the  spot visibility,  and  the  inhomogeneity
pattern must  be stable  over the time  interval when  the photometric
data  are collected.   Finally, the  non-periodic  phenomena mentioned
earlier  should not be  dominant contributors  to the  flux variation.
All these conditions are not always satisfied.

Most  of our  targets  are either  WTTS  or CTTS  which  show in  some
respects different patterns of variability.   In the case of WTTS, the
observed  variability is  dominated by  phenomena related  to magnetic
activity  which manifest themselves  on different  time scales,  as it
also occurs  in the more evolved  MS and post-MS  late-type stars (see
Messina  et al.   2004).  The  shortest time  scale, of  the  order of
seconds  to  minutes,  is  related  to  micro-flaring  activity.   Its
stochastic  nature  increases the  level  of  intrinsic  noise in  the
observed  time  series flux.   The  variability  on  time scales  from
several  hours to  days  is  mostly related  to  the star's  rotation.
Whereas,  the variabilities  on  longer time  scales,  from months  to
years, are related to the growth and decay of active regions (ARGD) as
well  as   to  the  presence   of  star-spot  cycles.   In   order  to
differentiate  the effects on  the variability  by ARGD  and star-spot
cycles from the effect of rotation, on which we are presently focused,
we have  analyzed our time series  data season by season  from 2004 to
2008. Excluding  cycle 5 which is  currently the longest  one, we have
collected our  data during observation seasons which  are shorter than
the timescale of ARGD typically observed in PMS stars.  In the case of
CTTS, apart from  cool spots, also hot spots  formed by accretion from
the disk  introduce additional variability.  This  type of variability
is quite  unexplored, in  the sense that  our knowledge either  of the
accretion processes  or of  their time  scales is not  as good  as for
WTTS.   We  know  from  previous   studies  that  in  most  cases  the
combination of different mechanisms operating on different time scales
makes the variability highly irregular.

We  included in  our analysis  also the  data collected  by  Herbst et
al. (2002) (hereafter referred to as H02), kindly made available to us
on our  request.  The  independent analysis of  the H02 data  with our
tools (described  in the following  subsections), allowed us  to check
the reliability  of our period  search procedures.  We  first searched
for the periodicity in our I  band time series data collected over all
the seasons, because the I band  data are more numerous than V as well
as R  band data (see  Fig. 2  and Table 1)  and also have  greater S/N
ratio.  Afterwards, the analysis of R  and V band time series data was
carried out  to either confirm  the periodicity found from  the I-band
data and/or to search for additional periodic variables.

 We have used the  Scargle-Press periodogram to search for significant
 periodicity. In  the following  sub-sections we briefly  describe our
 procedures to identify periodic variables among our targets.

\subsection{Scargle-Press periodogram}

The  Scargle technique  has  been  developed in  order  to search  for
significant  periodicities  in unevenly  sampled  data (Scargle  1982;
Horne \& Baliunas 1986). The algorithm calculates the normalized power
P$_N$($\omega$) for a given  angular frequency $\omega = 2\pi\nu$. The
highest   peaks  in  the   calculated  power   spectrum  (periodogram)
correspond to the candidate  periodicities in the analyzed time series
data.  In order  to determine the significance level  of any candidate
periodic signal, the height of the corresponding power peak is related
with a  false alarm probability (FAP),  which is a  probability that a
peak of given height is due to simply statistical variations, i.e.  to
noise.   This  method  assumes   that  each  observed  data  point  is
independent from the  others.  However, this is not  strictly true for
our time series data consisting of data consecutively collected within
the  same  night  and with  a  time  sampling  much shorter  than  the
timescales of periodic variability  we are looking for (P$^d$=0.1-20).
The impact  of this correlation  on the period determination  has been
highlighted by, e.g., Herbst  \& Wittenmyer (1996), S99, Rebull (2001)
and  Lamm et  al.   (2004).  In  order  to overcome  this problem,  we
decided  to determine  the  FAP  by different  ways  than proposed  by
Scargle (1982)  and Horne  \& Baliunas (1986),  the latter  being only
based  on the  number  of independent  frequencies.   We followed  the
method opted by H02 (hereafter referred to as \it Method A\rm) and the
approach  proposed by  Rebull  (2001) (hereafter  referred  to as  \it
Method B\rm) .

\subsubsection{Method A}
\begin{figure}
\begin{minipage}{8cm}
\centerline{\psfig{file=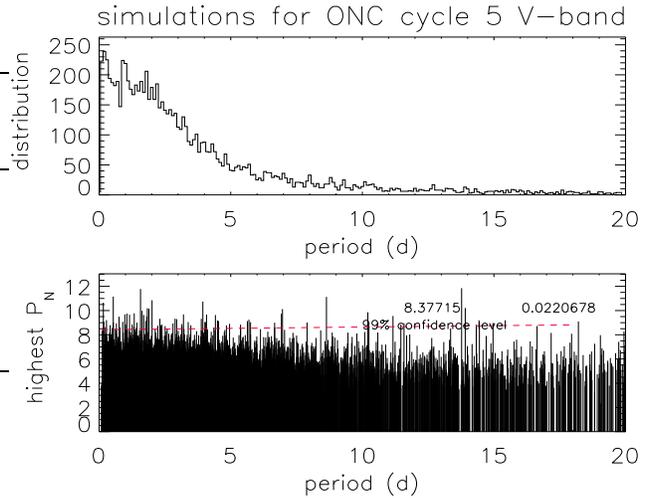,width=7cm,height=10cm,angle=90}}
\vspace{-0cm}
\caption{Distribution of periods (\it top panel \rm ) and power of the highest peaks (\it bottom panel \rm )  resulting from Scargle
 periodogram analysis of 10,000  ''randomised'' light curves. Original  time series data were collected during cycle 5 in the I band.}
\label{distri_permu}
\end{minipage}
\end{figure}

Following the  approach outlined by  H02, randomized time  series data
sets were created by randomly  scrambling the day number of the Julian
day (JD) while keeping photometric  magnitudes and the decimal part of
the JD unchanged. This method preserves any correlation that exists in
the original data set.  We noticed that Lamm et al. (2004) in order to
produce   the  simulated   light  curves,   randomized   the  observed
magnitudes,  instead of  the  epochs of  observation.  Then after,  we
applied the periodogram analysis to the "randomized'' time series data
for  a total  of about  10,000 simulations.   We retained  the highest
power peak and the  corresponding period of each computed periodogram.
In the top panel of Fig.\,\ref{distri_permu}, we plot the distribution
of detected periods  from our simulations for I- band  data of the 5th
cycle, whereas,  in the bottom panel  we plot the  distribution of the
highest power peaks vs. period.  The dashed line shows the power level
corresponding  to the  99\% confidence  level.  The  FAP related  to a
given  power P$_{ N}$  is taken  as the  fraction of  randomised light
curves that have  the highest power peak exceeding  P$_{ N}$ which, in
turn, is the  probability that a peak of this height  is simply due to
statistical  variations, i.e. white  noise.  We  note that  such power
thresholds are  different from season to season,  because of different
time sampling, length  of the observation season, and  total number of
data.   The normalised  powers corresponding  to  a FAP  $<$ 0.01  are
reported  in   the  second  column  of  Table\,\ref{fap}   for  the  5
observation seasons as well as the averaged H02 data.  We note that in
our analysis we  averaged the H02 time series data  in the same manner
as  done in  our  own  time series  (see  Sect.\,4.3). Therefore,  the
reduced total number of analysed data  in each light curve leaded to a
proportionally smaller  power level -  about 8.2 against 16.5  for the
unaveraged data as reported by H02.

\begin{table}
\caption{Normalized power at 1\% FAP as established by randomizing time series (Method A) and by correlated Gaussian noise (Method B)\label{fap}}
\begin{tabular}{c| c |r r r r r r}
\hline
cycle    & Method A & \multicolumn{6}{c}{Method B}\\
	& & \multicolumn{6}{c}{L$_{\rm corr}$(d)}\\
	&           &     0.0 & 0.1 & 0.25 & 0.5 & 1.0 & 2.0  \\	
\hline
	&  P$_{\rm N}$ & P$_{\rm N}$ & P$_{\rm N}$ & P$_{\rm N}$ & P$_{\rm N}$ & P$_{\rm N}$ & P$_{\rm N}$ \\
 1	& 6.2   	&	7.0 &	8.2 &	9.1	&	 9.3&  9.5	&	10.4	  \\
 2	& 6.9	  	& 7.8 & 	7.9	&	  8.3	&	9.2 & 9.9	&	10.3   \\
 3      & 6.6 	      	&  7.3 &   7.7 &   7.6&   	 7.7&  7.6 &	 8.6    \\
4	& 8.4		& 8.0  &    8.8  &  9.6 	&	10.3 &  11.4	&	13.4	   \\ 
 5	& 9.2 		& 8.4 & 	11.4&	 13.9& 	15.5	& 16.7	 &	19.6	  \\
H02  (averaged)   & 8.2 		& 7.8  & 	8.2 &  8.8 &	9.4  &  10.4 & 12.8 \\
\hline
\end{tabular}
\end{table}

\subsubsection{Method B}
Following  the method  given by  Brown et  al. (1996),  synthetic time
series data was generated with  the same sampling as the observed data
and the frequencies were searched over the same range as done with the
real  data. The  synthetic time  series are  built by  using  a random
number generator with a Gaussian distribution of points
\begin{equation}
x_{i >0} = \alpha x_{i-1} + \beta R(0,\sigma)
\end{equation}
where   $x$   represents   the   magnitude   of   the   light   curve,
$\alpha$=exp($-\Delta$  t/L$_{corr}$),  where  $\Delta$t is  the  time
between magnitudes  x$_{i-1}$ and x$_i$, $\beta$=(1$-\alpha^2)^{1/2}$,
and  R(0,$\sigma$) is the  random number  generator with  a dispersion
$\sigma$, which is  the variance of the time  series data. The initial
mean magnitude $x_0$ is selected via a call to R(0,$\sigma$). For more
details on this  procedure see Rebull (2001) and  Brown et al. (1996).
L$_{corr}=0$~d  sets the  case of  uncorrelated Gaussian  noise, where
each  data is  assumed to  be uncorrelated  from others.   A  range of
correlation  time starting from  0.1 to  2 days  were used.   For each
value of correlation time we have built a set of about 10000 synthetic
light  curves and  used the  distribution  of maximum  power from  the
corresponding periodograms to determine  the 1\% FAP.  The results are
summarized in Table\,\ref{fap}.

The  increase of L$_{corr}$  implies that  the synthetic  light curves
become more correlated and this makes the power level corresponding to
any threshold FAP larger, which in turn increases the risk to miss the
detection  of   real  periodicities.   We   found  that  a   value  of
L$_{corr}$=1.0d  is a  good  compromise to  properly  account for  the
correlation in our data.   For L$_{corr}>$1.0d we start systematically
missing even those rotation periods whose reliability have been firmly
established by previous multi-year surveys (e.g. H00).

We see that the power  level adopted to discriminate periodic from non
periodic target is  larger than the values found both  in the Method A
and  in Method  B for  uncorrelated Gaussian  noise (L$_{corr}=0.0$d).
Finally, we notice that these FAP values are conservative.  The reason
is that even in the same observing run the sampling of the time series
data differ  from object  to object.  This  happens primarily  due to,
objects close  to the edge of  the detector were some  time missed out
due  to telescope  pointing  error,  some time  at  poor seeing  faint
objects failed  to collect  sufficient signal and  hence undetectable.
When doing our simulations we  adopted the largest FAP level, which we
generally found  in the  most numerous time  series data (that  is the
case for more than 80\% of light curves).

 In order to establish whether a  star can be considered a periodic at
 99\% confidence  level we  decided to adopt  Method B  for correlated
 Gaussian   noise  (L$_{corr}=1.0$d).    All  the   stars   listed  in
 Table\,\ref{tab_results} satisfy this  selection criterion.  In a few
 cases, as  shown in  the on-line Figs.8-14,  more than one  peak were
 found with power  exceeding the fixed threshold value  at 99\%. These
 secondary  peaks are  generally  alias  of the  peak  related to  the
 rotation period,  which is  the only periodicity  we expect  in these
 late-type PMS stars in the period range searched by us (0.1-20 days).
 In order to check that the  secondary peaks are indeed nothing but an
 aliasing effect as well as to correctly identify the rotation period,
 we proceeded by  subtracting the smooth phased light  curve of period
 associated  with  highest  peak.   Then  after,  we  re-computed  the
 periodogram on the  pre-whitened time series data.  In  most cases no
 significant peaks were left in the pre-whitened time series data with
 confidence level larger  than 99\%.  A total 14  stars whose rotation
 periods were detected  with a confidence level larger  than 99\% were
 excluded   from  the   following   analysis  and   not  included   in
 Table\,\ref{tab_results}, since they  display very noisy phased light
 curves. These stars were also  not identified as a periodic variables
 by previous surveys.

\subsection{Uncertainty with the rotation period}

In order to compute the error associated with periods determined by us
we followed the  method used by Lamm et al.  (2004). According to this
method the uncertainty in the period can be written as
\begin{equation}
\Delta P = \frac{\delta \nu P^2}{2}
\end{equation}
where  $\delta\nu$ is  the finite  frequency resolution  of  the power
spectrum and  is equal to the full  width at half maximum  of the main
peak of the window function w($\nu$).  If the time sampling is not too
non-uniform,  which is  the  case of  our observations,  then
$\delta\nu  \simeq  1/T$,  where T  is  the  total  time span  of  the
observations.  From Eq.\,5  it is  clear that  the uncertainty  in the
determined period, not only  depend on the frequency resolution (total
time span)  but is also  proportional to the  square of the  power. We
also  computed the  error  on the  period  following the  prescription
suggested  by the  Horne  \& Baliunas  (1986)  which is  based on  the
formulation given by Kovacs (1981). We noticed that the uncertainty in
period computed according to Eq.\,5 was found to be factor 5-10 larger
than the  uncertainty computed by  the technique of Horne  \& Baliunas
(1986).  In this paper  we report  the error  computed with  the first
method  described above.   Hence, it  can  be considered  as an  upper
limit, and  the precision in the  period could be better  than that we
quote in this paper.

\section{Results}

\subsection{The H$\alpha$ line emission stars}

\begin{figure*}
\begin{center}
\includegraphics[scale = 0.8, trim = 00 235 0 00, clip]{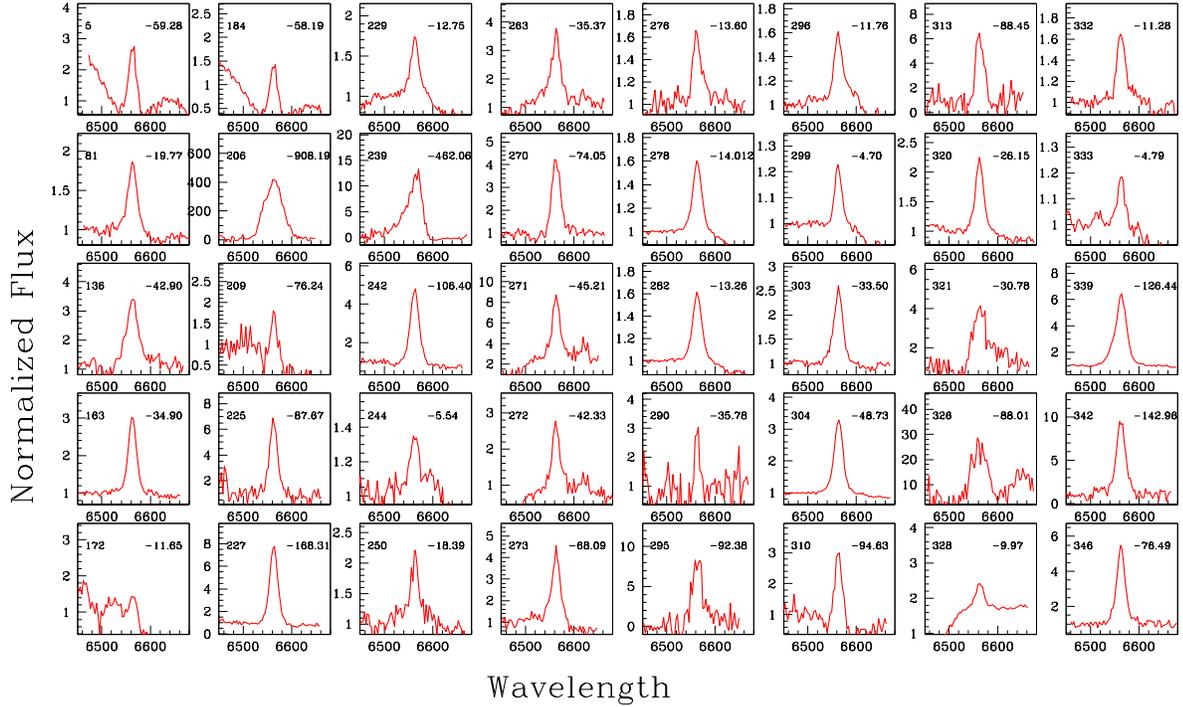}
\caption{ The spectra around H$\alpha$ from slit-less spectroscopic observations made on February 07, 2007 using  HFOSC.}
\label{onc_hctha}
\end{center}
\end{figure*}

The strong  H$\alpha$ emission  coming from low-mass  PMS stars  is an
unambiguous  signature of  active accretion  from the  disk on  to the
surface  of star.   Reliable knowledge  of the  presence/absence  of a
accretion disk is desirable particularly for studying evolution of the
angular momentum  in the  PMS phase. Moreover,  it helps us  to better
treat the light curves of  these stars which arise either from compact
hot-spots   and/or  from   the  variable   extinction   introduced  by
inhomogeneous  distribution of the  disk material.   Information about
H$\alpha$ emission of stars in our  field of view come either from old
objective prism surveys (Haro 1953; Wiramihardja et al.  1991) or from
the  recent surveys  made  by using  multi-object spectrographs  (S99;
Sicilia-Aguilar et al.  2005; Furesz  et al. 2008).  These surveys are
not complete. The old  objective prism surveys suffer from sensitivity
and  were unable to  obtain reliable  information about  the H$\alpha$
emission  from  faint  stars.   The  surveys  made  with  multi-object
spectrographs consist  of pointed observations  and, primarily, suffer
from selection  effect, i.e.  the  number of fibers were  limited and,
hence,  not  all the  ONC  PMS could  be  included  in these  surveys.
Therefore, we decided to do  a deep slit-less spectroscopy to identify
the  H$\alpha$ emitting  stars in  our ONC  field.  Details  about the
observations and the data reduction were given in Sect.~3.3.  From our
slit-less  spectroscopic  survey  we  could  identify a  total  of  40
emission lines stars with  moderate to very large H$\alpha$ equivalent
widths.   The EW  of these  emission  line stars  together with  other
relevant  informations are  given in  Table \ref{haew}.   Of  these 40
stars, 13 stars were not reported by the surveys mentioned above.  The
spectra around H$\alpha$ are  shown in Fig.~\ref{onc_hctha}.  The very
recent survey carried out by Furesz et al. (2008) has newly detected a
large number  of H$\alpha$  emitting stars. These  stars are  found to
have  CTTS-like  H$\alpha$  emission  with wide  wings  or  asymmetric
profile under the strong  nebular component.  Likewise objective prism
survey,  our slit-less  survey suffers  from the  problem of  high sky
background.   However, by  adding several  frames, doing  accurate sky
subtraction and,  finally, making use of  optimal spectral extraction,
we could obtain useful spectra of stars  as faint as 16 mag in I band,
and 18  mag in  V band. Interestingly,  we could not  detect H$\alpha$
emission  in very  few  cases,  although such  stars  were bright  and
previously reported as strong H$\alpha$ emitters.  We have also listed
such stars in Table \ref{haew}.   Altogether, 96 stars have been found
to be  H$\alpha$ emitting  stars in  our small FOV  and most  of these
H$\alpha$ emitting stars belong to low-mass CTTS group.
\begin{table}
\caption{\label{haew} The list of stars showing H$\alpha$ line in emission. We also present stars previously found to be strong H$\alpha$ emitter but not in our survey.}
\begin{center}
\begin{tabular}{|rrrrrrr|}
\hline
  \multicolumn{1}{c}{ID} &
  \multicolumn{1}{c}{JW} &
  \multicolumn{1}{c}{EW(\AA)} &
  \multicolumn{1}{c}{Period} &
  \multicolumn{1}{c}{Sky} &
  \multicolumn{1}{c}{I(mag)} &
  \multicolumn{1}{c}{Sp-Type} \\
\hline
5    & 265 &  -59.28   &  6.54  & SPN  &  14.26 & -        \\    
81   & 123 &  -19.77   &  9.61  & SC   &  12.85 & K2       \\
136  & 245 &  -42.90   &  8.82  & SPN  &  13.91 & M2e      \\
163  & 117 &  -34.90   &  9.01  & SC   &  13.17 & M0e      \\
172  & 288 &  -11.65   &  9.85  & SPN  &  13.02 & -        \\
184  & 272 &  -58.19   &  2.98  & SC   &  14.16 & M1.5e    \\
206  & 580 &  -908.19  &  ----  & SPN  &  16.56 & M1:e     \\
209  & 320 &  -76.24   &  ----  & SC   &  14.74 & K2       \\
225  & 138 &  -87.67   &  4.34  & SC   &  14.67 & M3.5e    \\
227  & 107 &  -168.31  &  1.07  & SC   &  14.13 & M1.5e    \\
229  & 239 &  -12.75   &  4.46  & SC   &  12.79 & M2.7     \\
239  & 277 &  -462.06  &  ----  & SC   &  15.29 & -        \\
242  & 235 &  -106.40  &  ----  & SC   &  13.77 & -        \\
244  & 576 &  -5.54    &  1.95  & SC   &  13.13 & M1.5     \\
250  & 135 &  -18.39   &  3.67  & SC   &  13.77 & M3e      \\
263  & 379 &  -35.37   &  5.59  & SC   &  15.19 & M5.2     \\
270  & 381 &  -74.05   &  7.75  & SC   &  13.96 & -        \\
271  & 632 &  -45.21   &  3.77  & SC   &  15.31 & M5.5     \\
272  & 628 &  -42.33   &  2.25  & SC   &  14.23 & -        \\
273  & 647 &  -68.09   &  8.20  & SC   &  13.36 & M5e      \\
276  &  91 &  -13.60   &  16.67 & SC   &  13.49 & M4       \\
278  & 416 &  -14.01   &  2.11  & SC   &  14.16 & M3.5     \\
282  & 421 &  -13.26   &  8.84  & SC   &  11.67 & G7:e     \\
290  & 294 &  -35.78   &  2.57  & SC   &  15.41 & M4.5     \\
295  & 715 &  -92.38   &  ----  & SC   &  15.21 & M5.5     \\
296  & 673 &  -11.76   &  3.22  & SC   &  13.11 & M5       \\
299  & 165 &  -4.70    &  5.74  & SC   &  11.85 & A7       \\
303  & 101 &  -33.50   &  1.05  & SC   &  14.42 & G:e      \\
304  & 328 &  -48.73   &  ----  & SC   &  12.80 & K4       \\
310  & 518 &  -94.63   &  ----  & SC   &  15.00 & -        \\
313  & 649 &  -88.45   &  1.80  & SC   &  15.69 & M6       \\
320  & 313 &  -26.15   &  ----  & SC   &  13.40 & M0       \\
321  & 447 &  -30.78   &  2.60  & SC   &  15.17 & M4       \\
326  &5159 &  -88.01   &  2.39  & SC   &  17.07 & -        \\
328  & 498 &  -9.97    &  7.27  & SC   &  13.95 & -        \\
332  & 225 &  -11.28   &  ----  & SC   &  13.60 & M1.5     \\
333  & 501 &  -4.79    &  9.69  & SC   &  13.14 & M0       \\
339  & 295 &  -126.44  &  2.85  & SC   &  13.15 & M2       \\
342  &  73 &  -142.96  &  2.23  & SC   &  14.45 & M1e      \\
346  & 284 &  -76.49   &  3.08  & SC   &  14.17 & M3e      \\
\hline
     &  &           &   No H$\alpha$     &      &        &          \\
\hline
18   & 278 &  -59.0    &  6.84  &SPN   &  13.84 & K4:e     \\
45   & 127 &  -43.0    &  ----  &SC    &  13.9  & -        \\
105  & 192 &  -25.0    &  8.93  &SC    &  13.77 & M2       \\
148  & 258 &  -20.0    &  9.94  &SPN   &  13.8  & M0.5     \\
204  & 334 &  -65.0    &  5.34  &SC    &  14.27 & -        \\
216  & 422 &  -22.0    &  5.94  &SC    &  14.46 & -        \\
327  & 115 &  -86.0    &  ----  &SC    &  15.99 & M5.5     \\
\hline                                               
\end{tabular}

\end{center}
\end{table}

\subsection{The color-magnitude diagram}
\begin{figure*}
\begin{center}
\includegraphics[scale = 0.60, trim = 00 00 00 00, clip]{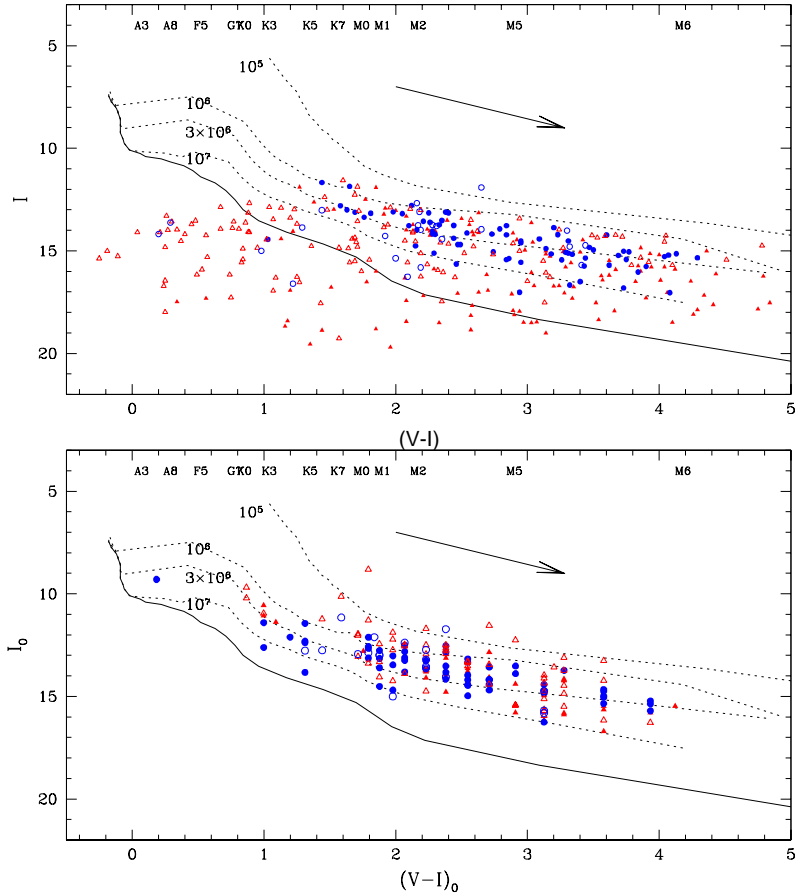}
\caption{The   color-magnitude   diagram of ONC stars  without reddening correction is shown in the top panel. The bottom panel shows the same diagram after reddening correction. Filled and open data points represent stars inside or outside nebulosity, respectively. The circles represent CTTS and  other stars are marked by a triangle.}
\label{onc_hri}
\end{center}
\end{figure*}

The  color-magnitude diagram  (HR diagram)  is the  most  reliable and
widely used  tool to obtain mass,  age, and radius  of cluster members.
These fundamental  stellar parameters are determined  by comparing the
position of  stars on HR  diagram with pre-main  sequence evolutionary
tracks.  Furthermore,  the distribution of stars  with respect to PMS isochrones may
 reflect  the star formation history  (Lada \& Lada  2003).  It is
also an  effective tool  to explore the  evolution of  stellar angular
momentum in the  PMS phase.  To construct the HR  diagram we need very
accurate measurements of effective temperatures (colors) as well as of
 luminosities  (magnitudes)   of  stars.   The  precise  effective
temperature  is  usually  measured  from  spectroscopy,  whereas,  the
luminosity  is  obtained  from   de-reddened  as  well  as  bolometric
corrected  magnitudes.    The spectral coverage and resolution of our slit-less 
spectra does not allow us to obtain accurate spectral type,  and hence as the  temperature, or
equivalently, the intrinsic color  is concern we rely on Hillenbrand
(1997)  data. On  the  other hand,  much  more accurate  I  and V  band
magnitudes are  determined from  our own long-term  observations.  The
median value determined  from the data of all  five cycles is expected
to  be less affected  by the  astrophysical error  (Hillenbrand 1997).
The   I  vs.   V$-$I   color-magnitude  diagrams,   before  and   after
inter-stellar extinction  correction, are given  in Fig.~\ref{onc_hri}.
The A$_V$ and  (V-I)$_o$, which are taken from  Hillenbrand (1997) are
available for only 50\% of our program stars. Therefore, the reddening
corrected CMD could be only made for a sub-sample of our data. We have
also over-plotted  the ZAMS  and various isochrones  from Siess  et al.
(2000). The solar metallicity with no convection overshooting was used
to  compute the  ZAMS and  the isochrones.   The  theoretical effective
temperatures and luminosities were  converted into (V$-$I) color and V
and I magnitudes  by making use of the conversion  tables of Kenyon \&
Hartmann (1995).

 We found a large number of faint blue objects in the HR diagram lying
 below the ZAMS.   Because the reddening vector is  almost parallel to
 ZAMS as  well as the  isochrones, so even after  reddening correction
 these stars will not fit to any of the isochrones. Hillenbrand (1997)
 also found such blue and less-luminous objects and inferred that they
 may be either heavily veiled  CTTS and/or stars buried in nebulosity.
 Strongly veiled stars  can become systematically bluer by  1-2 mag in
 (V$-$I) color.  However, at the  same time they should also be strong
 H$\alpha$  emitters,  which  is  indicative  of  strong  active  disk
 accretion, that  has not been found  except for few  objects.  On the
 other hand, there is a large number of such blue faint stars lying in
 the intense  nebulosity.  Since the  average color of the  ONC nebula
 close  to the  Trapezium is  about V$-$I=0.11  mag, therefore,  it is
 likely that  these stars appear  bluer due to  nebular contamination.
 These  two phenomena  can not  explain all  the  outliers.  Moreover,
 there are  a few  faint blue non-accreting  stars outside  the strong
 nebulosity.  These objects may not  be at all stellar objects or they
 may be  UX Ori objects, which have  been found to be  bluer when they
 become fainter.  Most of these  faint blue objects were  not observed
 spectroscopically  by  Hillenbrand   (1997)  and  hence  they  simply
 disappear in the  reddening corrected CMD.  It also  appears from the
 uncorrected CMD that  CTTS are more populous in  the upper portion of
 the PMS sequence  than WTTS.  This indicates that  accreting CTTS are
 in general younger  than WTTS.  But any such  trend disappears in the
 reddening corrected CMD.  In agreement with earlier studies, the mean
 age of ONC appears  to be of about 1~Myr with an  age spread of about
 3~Myr in  the uncorrected  CMD.  Here, we  assume that  the reddening
 vector is almost parallel to the isochrones in the low-mass range and
 that  the   reddening  correction   will  not  change   the  relative
 distribution  of the  stars.   In the  reddening  corrected CMD,  ONC
 appears  slightly older ($\sim$2~Myr).   This apparent  difference in
 the age  could be  due to incompleteness  of the  reddening corrected
 sub-sample.

\begin{table*}
\centering
\begin{minipage}{\linewidth}
\caption{Result of the periodogram analysis of periodic variables of the ONC.\label{tab_results}}
\scriptsize
\begin{tabular}{lrrrrrrrrrrrrrrr}
\hline\hline
ID &JW  &  Power  &  P$\pm\Delta$P     &   $\chi^2_{\nu}$  &  $<\sigma>$  &  $\Delta$I   &  \# &  \# & \#  &  n1  &   n2  &  n3   &  Sky  &   Object~Type &   Neighbour \\
 &     &         &   (d)    &    &  (mag)  & (mag) &  obs. &  mean & dis. &   &   &   & &  &\\
\hline
  5  &265   &          45.59  &    6.540 $\pm$  0.100 &  73.17 & 0.041  &    0.34  & 230  &  61  &   2  &   c4  &        c3/c5/H  &     new &   SPN &  C &   y \\
 15  &710   &          15.68  &    7.810 $\pm$  0.180 &    106.99   &  0.231  &    1.43  &  76  &  25  &   2  &   c2  &        c1/H  &      =S &    SC &  C &   - \\
 16  &349   &          50.41  &    9.250 $\pm$  0.120  &    173.62   &  0.054  &    0.63  & 397  & 101  &   6  &   c5  &          c3/c4  &     new &   SNB &  - &   - \\
 17  &125   &          16.71  &    8.860 $\pm$  0.150  &     86.95   &  0.010  &    0.11  &  74  &  21  &   2  &    c3  &          c1/c4  &     new &    SC &  C &   - \\
 18  &278   &          60.60  &    6.840 $\pm$  0.060  &   4189.88   &  0.016  &    1.60  & 400  & 102  &   6  &   c5  &            all  &      =H &   SPN &  C &   - \\
 23$^{\rm a}$  &366   &18.86  &    8.790 $\pm$  0.230  &     36.30   &  0.026  &    0.16  &  97  &  29  &   1  &   c2  &              -  &     =H &   SNB &  - &   - \\
 27  &437   &          31.38  &    2.341 $\pm$  0.008  &     11.60   &  0.018  &    0.09  & 180  &  88  &   5  &   c5  &              H  &   =H   &   SNB &  C &   - \\
 29  &417   &          15.58  &    7.370 $\pm$  0.100  &      7.64   &  0.035  &    0.08  &  75  &  21  &   1  &   c3  &             c1/H  &     =H &   SNB &  C &   - \\
 31  &622   &          19.93  &    3.770 $\pm$  0.150   &     52.29   &  0.032  &    0.26  & 365  & 105  &   4  &   c5  &           c1/H  &     new &   SNB &  - &   - \\
 34  &81    &          25.05  &    4.400 $\pm$  0.040  &     26.53   &  0.010  &    0.12  &  62  &  21  &   2  &   c4  &     c1/c2/c3/H  &      =H &    SC &  W &   - \\
 35  &9213  &          21.64  &   12.220 $\pm$  1.410  &     16.13   &  0.039  &    0.34  & 103  &  21  &   1  &   c1  &             c3  &     new &   SNB &  - &   y \\
 40  &317   &            6.33  &    8.080 $\pm$  0.190  &     14.37   &  0.018  &    0.08  &  97  &  29  &   3  &   c5  &          c2  &      =H &   SNB &  - &   - \\
\hline
\multicolumn{16}{l}{NOTE:  Only  a portion of the table is shown here and complete table is  available only in  electronic edition of the MNRAS.}\\
\multicolumn{16}{l}{a: The rotation period was detected in only one season and, although with a  FAP $<$ 1\%, it needs to be confirmed by future observations.}\\
\multicolumn{16}{l}{b: The rotation period, although detected in multiple seasons and with a  FAP $<$ 1\%,  may be  a beat period, being very close to the window function main peak.}
 \end{tabular}
\end{minipage}
\end{table*}

\subsection{Periodic variables}

\begin{figure*}
\includegraphics[width=10cm,height=16cm,angle=90]{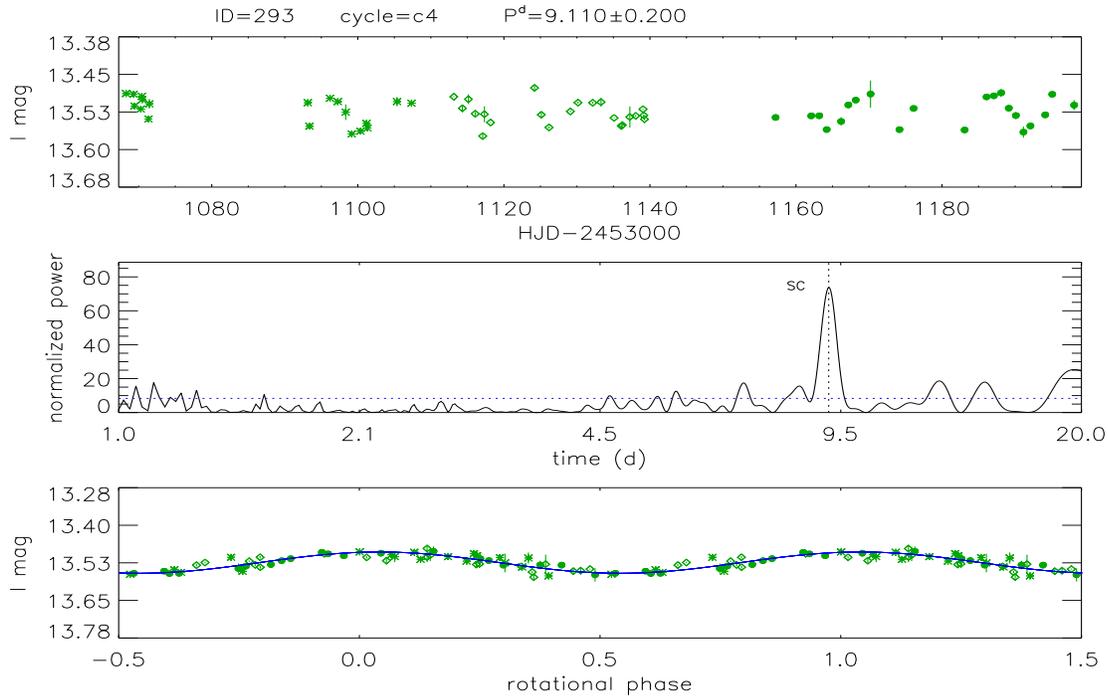}
\caption{Results of periodogram analyses on the newly discovered periodic variable ID=293. \it Top panel: \rm I-band  time series data from  cycle 4. Different symbols are used to 
better distinguish three different time intervals within the same observations season. \it Middle panels : Power spectra from Scargle analysis.
The horizontal dashed line indicates the 99\% confidence level, whereas the vertical dotted line marks the detected periodicity. \it Bottom panel  \rm phased light curve 
using the detected period.}
\label{example_lc}
\end{figure*}

The results of our search for periodic variables are presented in this
subsection.   As  mentioned,  we  searched  for  rotation  periods  by
analyzing  our  own data  collected  in  five consecutive  observation
seasons as  well as H02 data.  Only  stars having a peak  power in the
periodogram, larger  than 99\% confidence level  computed according to
Method B  for correlated Gaussian  noise, were selected as  a periodic
variables. (see Sect.   4.6.2).  In Table\,\ref{tab_results} we report
the  information related  to only  the  season in  which the  rotation
period has  been determined most  precisely.  Table\,\ref{tab_results}
lists the  following information: our star  identification number (ID)
which runs  from 1 to  346; an identification number  from Hillenbrand
(JW number);  the normalized power (P$_{  N}$) of the  highest peak in
the Scargle power spectrum, and  the rotation period together with its
uncertainty (P$\pm\Delta$P).  In the  next columns we list the reduced
chi-squares ($\chi^2_{\nu}$) of the  light curve computed with respect
to   the  median   seasonal  magnitude   and  the   average  precision
($<\sigma>$)  of the  time series  data computed  as described  in the
Sect.\,4.3.  Then we list the amplitude  of the light curves in I band
($\Delta$I), which  was computed by making the  difference between the
median values of  the upper and lower 15\% of  magnitude values of the
light   curve  (see,   e.g.,  H02).    That  allows   us   to  prevent
overestimation of the amplitude due to possible outliers.  After this,
we list  the number of total  useful observations and  the data points
after averaging the close observations.   In the next three columns of
Table\,\ref{tab_results} we  put the  following notes: n1  denotes the
season to which  the listed period and all the  values in the previous
columns  refer; n2  denotes  the cycles  where  the same  periodicity,
approximately within the computed  uncertainty, were found ('H' stands
for H02 and H00 data, 'S' for S99 data, 'all' for all cycles including
H02,  H00 \&  S99 data);  n3  indicates whether  the star  is a  'new'
periodic or  a previously known  periodic variable whose period  is in
agreement or  disagreement with the  one determined by Stassun  (S) or
Herbst  (H). Then  after the  position of  stars with  respect  to the
nebulosity  (Sky),  the  star  classification  as  CTTS  (C)  or  WTTS
(W). Finally, the last column denotes the presence (y) of another star
closer  than 6  arc-sec.   In Fig.\,\ref{example_lc}  we  plot, as  an
example, the  result of our  periodogram analysis obtained for  one of
our  targets ID=293 which  has been  identified as  periodic variable.


\begin{figure}
\includegraphics[width=8cm,angle=0]{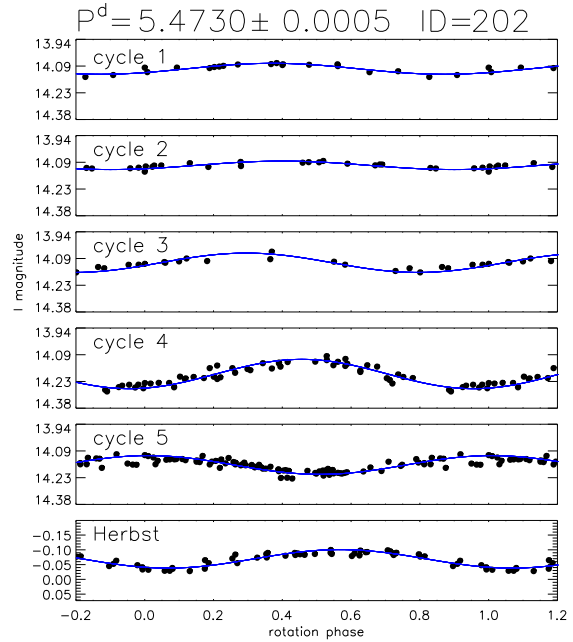}
\vspace{-1cm}
\caption{Example of light curves of one of our CTTS targets (ID=202) whose rotation period was detected in all our 5 observation season and in the Herbst et al. (2002) data as well.}
\label{target_202}
\end{figure}

As  listed in  Table\,\ref{tab_results}, there  is a  large  number of
stars whose rotation period has been detected in all observing seasons
including H02 data as well.  The periodogram analysis performed on the
whole 5-yr  time series  allows us to  determine the  rotation periods
with accuracy  better than 1\%.   Although the study of  the long-term
behaviour of  our targets  will be the  prime subject of  a subsequent
paper, we  show in Fig.\,\ref{target_202}  the light curves of  one of
these stars (ID=202),  as an example.  This example  light curves show
two  interesting  features.  The  first  is  that  within the  longest
observing run (season 5), the light curve is very smooth and the shape
has remained  constant over about  5 months.  This indicates  that the
size as well as the distribution  of the spot which modulates the star
light remain unchanged.  This  behaviour differs from what is observed
in MS stars of similar  rotation period, whose spot patterns are found
to be stable  over periods not longer than 1 to  2 months. However, we
stress that  this is  the case for  a limited  number of stars  in our
sample,  whereas there  is  a  large number  of  stars whose  rotation
period, although detected  in cycle 5 data, was  not detected when the
whole time  series data was analysed.   This is likely due  to a rapid
change of  the spot  pattern on these  stars.  Another feature  is the
change in amplitude, mean light level and migration of the light curve
minimum from season to season.  These latter variations may be related
to the the presence of surface differential rotation (SDR) and ARGD.

 As   already  mentioned,   a  part   of  our   ONC  field   has  been
 photometrically  monitored since  1990  at Van  Vleck Observatory  by
 Herbst and  collaborators (see,  e.g., H00), in  1994 by S99,  and in
 1998-1999 by H02 at the MPG/ESO 2.2m telescope.  Our analysis allowed
 us to identify a total of 148 periodic variables: 56 are the periodic
 stars newly discovered by us,  whereas 92 were already known from the
 mentioned  previous surveys  (24  from Herbst  and collaborators,  11
 stars from  S99, and  57 from  both groups).  We  did not  detect any
 periodicity  in 18  stars, of  which  2 were  previously reported  as
 periodic variables  only by  S99 (star  302 and 342)  and 16  only by
 Herbst and  collaborators (star 3, 4,  38, 39, 62, 69,  94, 114, 129,
 152,  175, 208,  302, 327,  339, and  344).  Actually,  we  note that
 previous surveys could detect the  periodicity for 14 out of these 18
 stars in only one observation season, hence their classification as a
 periodic variables  can be considred  tentative.  Among the  92 stars
 with already known  rotation period, there are 14  stars for which we
 detect a period different  than that previously reported. However, we
 could understand the  origin of the disagreement.  For  5 stars (star
 81, 96, 135,  181, 307), the rotation periods  reported by either H02
 or S99 are very different  from ours.  Looking at our periodograms of
 H02 averaged  data we  found a  major power peak  at the  same period
 found  in  multiple  seasons  in   the  power  spectra  of  our  time
 series. However,  such peaks have  a power slightly smaller  than the
 threshold (P$_N$=16.5) that they  fixed and, therefore, they were not
 considered  in the H02  sample.  Indeed,  this is  an example  of the
 advantage of multi-season  with respect to single-season observations
 to obtain more reliable rotation periods.

For the stars  240, 254, 283, 296  we found a period  that is
almost doubled  from the period  reported by H02 or  S99.  This
mismatch can be  explained by considering  that, at that time
when these stars  were observed, two major spots
were located  on opposite stellar  hemispheres, giving rise to  a flux
modulation with half the period we found.

The  disagreement to 4 stars (star 44,
285,  293, 303) is  likely due  to the  beat phenomenon,  ours  or the
H02 and S99  periods being beats  of the true  period.  For  star
274 we found  in all seasons two different periods,  one in common with
H02 \& S99 and another different. We adopted the latter, since it gives less
scattered phased light curves.

Therefore, we can state that for 92 periodic stars, the results of our
period search  well conciliate with  the results of  previous studies.
This  point is  very important,  since it  assures that  we  have been
indeed collecting  very precise  photometric data as  well as  using a
reliable  period  analysis technique.   In  addition  to  this we  get
confidence  on  our  new  detection  of periodic  variables  and  also
confirmation  of the  previous findings.   We have  discovered  56 new
periodic  variables in  our very  small 10$\times$10  arc-min  FOV and
hence increased the total number of known periodic variables by almost
50\%.  However,  we must  note that 7  new periodic  stars (identified
with  an  apex  'a'  in  Table\,\ref{tab_results})  were  detected  as
periodic for the first time by us  and in only in one season.  We need
additional observations to confirm  these periodicities, which will be
considered at the present as tentative and will not be included in any
subsequent  analysis  or  statistics.   For instance,  the  number  of
periodic  stars whose  period is  detected in  only one  season ($\sim
5\%$) is similar to the total number of false detections expected in 5
observation seasons by  fixing the FAP at 1\%.   We must consider with
great caution  also the 2 new  periodic stars identified  with an apex
'b' in Table\,\ref{tab_results}, since their period is very close to 1
day. Although, the same period  was detected in multiple seasons, with
high confidence  level and  giving smooth light  curves, it may  be an
alias arising  from the observation  day-night duty cycle.   They also
will not be considered in the following analysis.  The complete sample
of   light  curves   of  our   periodic  variables   are   plotted  in
Figs.\,\ref{curve1}-\ref{curve5}.  In  these figures we  plot only one
light  curve per  star, specifically  the  details of  which has  been
listed in Table\,\ref{tab_results}.

 There  are very few  massive early  spectral type  stars in  our FoV.
 Since such  objects are saturated,  almost all our program  stars are
 low- to very  low-mass objects and we treat them as  TTS. In our FOV,
 till  to date,  96 stars  are  found to  show the  H$\alpha$ line  in
 emission and with very large EW, hence, these stars can be assumed as
 accreting  CTTS.  Whereas,  the remaining  242 stars  which  lack the
 H$\alpha$ emission can  be considered as WTTS. These  numbers of CTTS
 and  WTTS should  be considered  a preliminary  estimate,  and likely
 subjected  to be changed  in the  future.  In  fact, as  mentioned in
 Sect.  5., a few  stars which  were reported  by previous  studies as
 H$\alpha$ emission stars,  are found by us with  no emission, whereas
 13 newly H$\alpha$  emitting stars have been identified  by us. Thus,
 it  appears  that to  identify  all  accreting  CTTS using  H$\alpha$
 emission as  a proxy, more extensive  multi-year spectroscopic survey
 is needed.  Moreover, we note that a few stars (like star 16, 41, 297
 and 309)  show an I-band light  curve amplitude larger  than 0.5 mag,
 although  we  have  not   detected  H$\alpha$  emission.  Such  large
 amplitude most likely arises from  the presence of both hot spots and
 variable  extinction  by  the  circum-stellar  disk.   Keeping  these
 numbers distribution of  WTTS and CTTS, our period  search shows that
 about 72\% of CTTS are periodic variables. Whereas, the percentage of
 periodic  WTTS is about  32\%.  Our  findind is  contrary to  Lamm et
 al. (2004), whose results on photometric monitoring of NGC2264 reveal
 that 85\% WTTS and 15\% CTTS are periodic variables.

 From  our study  it appears  that  the detection  of periodic  signal
 related  to  rotation  is  easier  in  CTTS than  in  WTTS.  This  is
 surprising in the sense that, as mentioned previously, it is believed
 that the variability of  CTTS is dominated by non-periodic phenomena,
 which  should often  prevent  the detection  of  any periodic  signal
 coming  from the rotation  (Herbst et  al. 1994;  Lamm et  al. 2004).
 From the present study, it appears that the CTTS of ONC have patterns
 of  surface inhomogeneities which  give rise  to the  flux variations
 more stable  than in WTTS and much  more than in MS  stars of similar
 period  (see Fig.\,\ref{target_202}  as  one example  of stable  CTTS
 light curve).   The increased ability  to detect rotation  periods in
 CTTS  may  be due  to  the large  amplitude  of  light modulation,  a
 characteristic of CTTS.  That compensates to some extent the presence
 of "noisy" non-periodic  phenomena, allowing the periodogram analysis
 to correctly  identify the rotation period. The  other possibility is
 that the  number of CTTS  identified from various  H$\alpha$ emission
 surveys  has  been underestimated,  and  actual  number of  H$\alpha$
 emitting CTTS  are larger  than the  total number of 96  what we used
 for the calculation.

As  we mentioned  earlier, we  have divided  our entire  FOV  in three
different regions and  the number of spatial distribution  of total as
well  as  periodic variables  are  as  follows.   There are  59  stars
residing in bright nebular region  and designated as SNB, 98 stars are
in partially  nebular region (SPN),  189 are outside the  nebula (SC).
We could  determine the rotation  period of 27\% (16/59)  stars inside
the  nebula, of  33\% (33/98)  partially in  the nebula,  and  of 52\%
(99/189) outside  the nebula.  As expected, the  detection of periodic
variables outside nebulosity,  where the background light contribution
is minimum  and the photometric  accuracy better, is much  higher than
inside  nebulosity.  Another possibility  could be  the presence  of a
large number of intermediate mass  stars, inside the nebula, which are
known for their irregular photometric behaviour.  Finally, it was also
found from  previous studies that low-mass stars  inside the Trapezium
cluster are relatively younger (Hillenbrand 1997).  Hence, it is quite
possible that  the youthness of  these objects make  their variability
rather irregular.

\begin{figure*}
\begin{center}
\epsfig{file=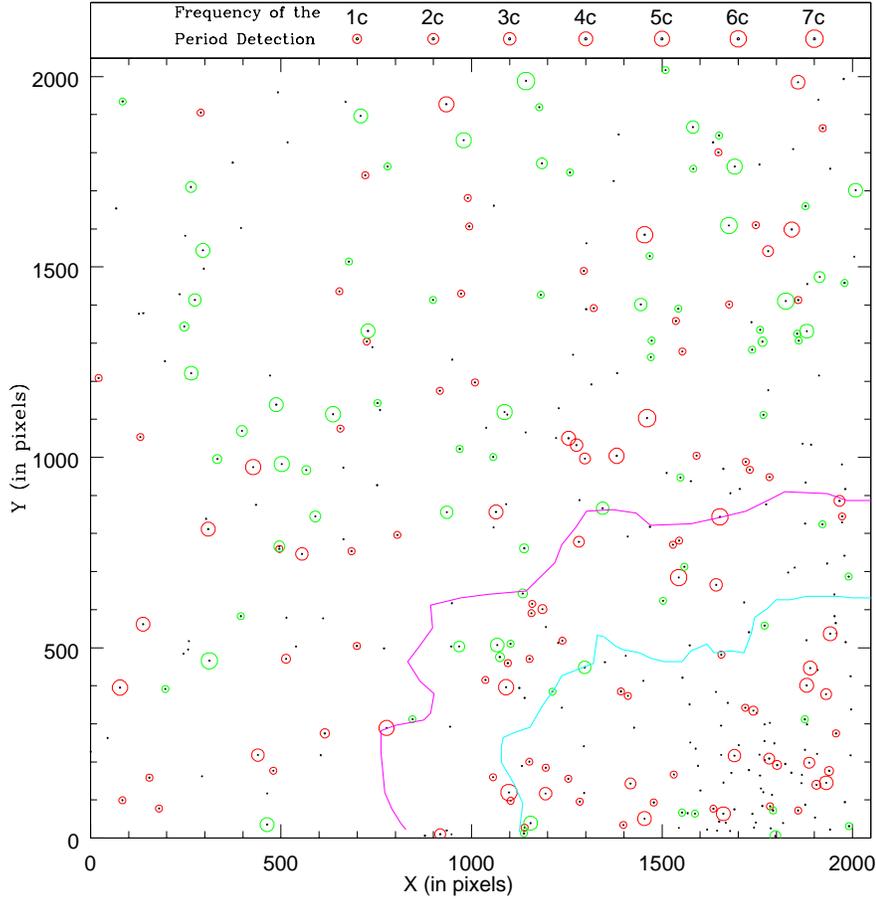,scale=0.60,angle=0}
\caption{The   spatial   distribution of periodic variables. Open circle represent the periodic variable detected from our long-term monitoring program. Whereas the  detection frequency is represented by size  of the open circle, i.e. the largest circle correspond to the  periodic variables, detected in all cycles including the previous surveys by Herbst et al. (2002) and Stassun et al. (1999). The green color represent the CTTS stars identified from \ha emission and the red color represent WTTS and/or other unclassified stars found to be periodic. }
\label{onc_period_sdist}
\end{center}
\end{figure*}

The part  of our FOV is very  much affected by the  presence of strong
HII nebular  emission and, hence, the background  contribution is very
high. In addition to this as mentioned earlier the distribution of the
stellar mass as well as age are not uniform in this region, therefore,
we attempted to get some information about the spatial distribution of
periodic  variables.   In  Fig.\,\ref{onc_period_sdist}  we  plot  the
spatial  distribution  of  periodic   variables  over  the  FoV.   The
unexpected result what we find  is that the detection frequency of the
rotation period, i.e.  number of  seasons on which any object has been
identified as periodic, is indistinguishable from inside to outside of
the nebula.   This result is  surprising because the fact  that inside
the  nebula, the  observed magnitudes  of stars  are subjected  to the
lower photometric precision, objects  are relatively massive and their
age relative to the mean cluster age is less.

\subsection{Scarcity of short period variables}

During cycles 4  and 5, we dedicated a number of  nights to search for
very  fast rotating  periodic variables.   In order  to make  our time
series  data sensitive  to very  short-term flux  modulations  down to
about  0.1  days,  we  monitored  the  ONC  field  over  several  full
nights. Again,  the data collected  within one hour were  combined and
used as one single data point while doing time series analysis.  These
more numerous data collected within  the same night allowed us to very
accurately  identify 22  periodic variables  with period  shorter than
about two days, 5 of which  were newly identified by us. We could also
identify 2 new variables having  periods very close to one day.  These
short period  variables are reported  in this paper because  they have
been  detected  in  multiple   seasons  with  high  confidence  level.
However, their  period need  to be confirmed  since they may  be alias
arising from  the observation day-night  duty cycle.  The  scarcity of
very  fast rotators  (P$<$0.9d) in  the magnitude  range (12$<$I$<$17)
explored  by  our survey  supports  the  S99  finding on  short-period
variables.  He  first noticed that  the ultra fast rotators  that have
been found in the Pleiades or in other older clusters are quite absent
in the ONC, at least in the 12$<$I$<$17 magnitude range.  This finding
is  quite expected  for the  low-mass stars  of 1~Myr,  whose break-up
velocity is  close to  0.5 days.  However,  here we like  to emphasize
that the  fastest rotating stars  identified by us  in our FOV  show a
period  about two times  longer than  S99.  In  contrast, after  a few
Myrs, like in  NGC\,2264, the number of rotators  faster than 0.4 days
increases  rapidly.   A  very  deep  photometric  monitoring  campaign
recently carried  out by Rodriguez-Ledesma  et al. (2008) in  the ONC,
shows the  presence of  very fast rotators  (P$>$0.5d) among  Very Low
Mass  Stars  (VLM) and  sub-stellar  objects  (BDs).  These very  fast
rotating  VLM and  BDs  probably possess  different dynamo  mechanism,
operating  in  this  mass  regime  and producing  magnetic  fields  of
different strength and topology as well (Herbst et al. 2006).
\begin{figure*}
\epsfig{file=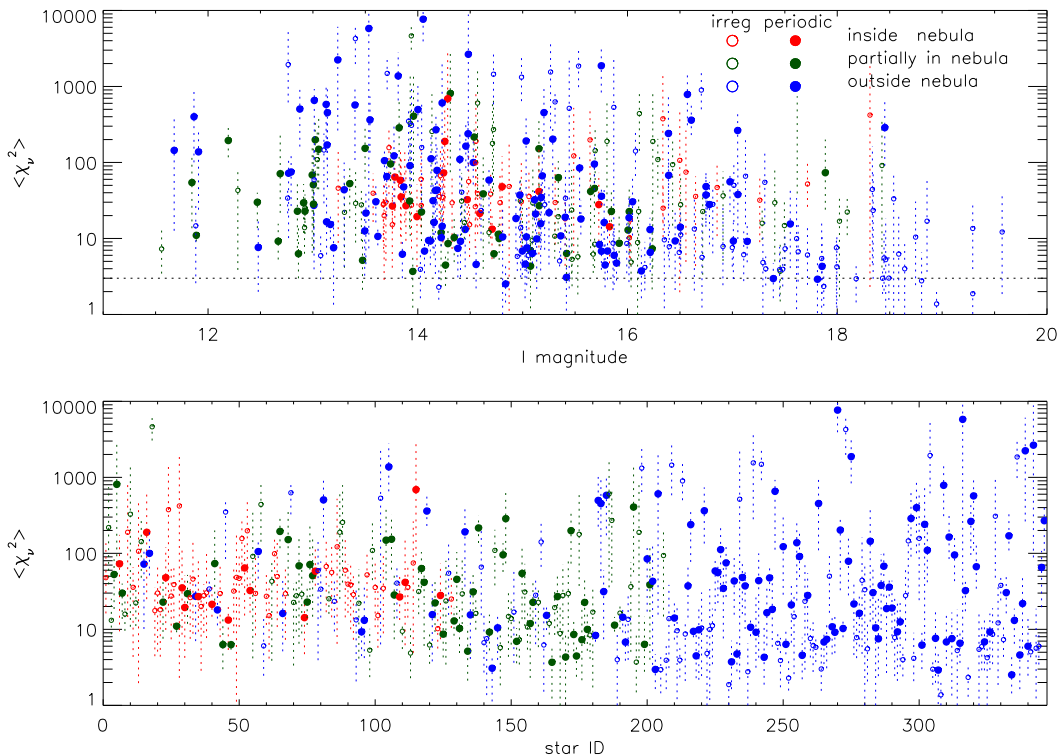,scale=0.65,angle=90}
\caption{\it Top panel: \rm Distribution of reduced chi-squares vs. I magnitude. Filled and open symbols are used to distinguish periodic from non periodic variables, whereas different color indicate the star's position with respect to the nebula. \it Bottom panel: \rm reduced chi-squares vs. internal ID number.}
\label{chis}
\end{figure*}
\subsection{Non-periodic targets}

 We never detected evidence of  periodic variability in 198 out of 346
 stars.   We have  computed  for each  season  the reduced  chi-square
 ($\chi^2_{\nu}$) with  respect to the seasonal median  light level in
 order to quantify their  level of variability.  The $\chi^2_{\nu}$ of
 each star is  found to vary from season to  season. Their mean values
 vs.   the median I  magnitude (top  panel) and  vs.  our  internal ID
 number (bottom  panel) are  plotted in Fig.\,\ref{chis}.   The dashed
 vertical bars  connect the minimum and  maximum $\chi^2_{\nu}$ values
 ever observed.   The red, green  and blue colors indicate  the target
 position  with respect to  the nebula  (inside, partially  inside and
 outside, respectively).   Filled and open  bullets represent periodic
 and non periodic variables, respectively.

Except the  stars 230  and 308,  which are the  least variable  in our
sample, all non periodic targets can be classified as variable, having
$\chi^2_{\nu}>3$ in  at least  one season.  This  is not  a surprising
finding since  all targets  are very young  stars. They  host numerous
phenomena which induce variability in  the optical as well as in other
bands.  Interestingly,  the $\chi^2_{\nu}$ show  a spatial dependence:
their values monotonically decrease from inside to outside the nebula.
This effect does not depend on  the star's brightness, as shown in the
top panel  of Fig.\,\ref{chis}, but more  likely on mass  and age.  As
mentioned, a  previous study (Hillenbrand 1997)  shows a concentration
of intermediate-mass stars inside  the nebula, which are characterized
by very  irregular and large-amplitude variability.  It  is also quite
possible that  stars inside nebula  are younger and,  therefore, being
more active give rise to a larger-amplitude variability.

\section{Discussion and  plan for near future}

  The  luminosity  spread in  the  color-magnitude  diagram for  stars
  having similar color, and  hence mass, has been generally associated
  with the age spread and found to be an effective tool to explore the
  star formation history in young clusters (Hillenbrand 1997; Palla \&
  Stahler   2000;  Hartman   2001).    However,  limited   photometric
  measurement accuracy,  intrinsic variability and  effect of binarity
  in the magnitudes can mask the signatures of the initial age spread.
  The  use of  the  median magnitude  from  our long-term  photometric
  monitoring of the  ONC stars has allowed us to  construct a CMD much
  more  accurate than in  previous studies,  mostly based  on snapshot
  observation,  and,  hence,  has  allowed  us  to  better  asses  the
  presence/absence of any age spread  in the ONC. It is quite apparent
  in the  CMD of Fig.~\ref{onc_hri} that,  after reddening correction,
  both WTTS  and CTTS  indeed show  a spread in  the magnitude  at any
  color, which possibly  points toward the presence of  an age spread.
  However, we can  not draw a very firm conclusion  on this because of
  the small size of our  sample and possible effects from unrecognized
  binaries   as   well  as   uncertain   extinction  correction.   The
  presence/absence of spread in age of young stars has been subject of
  exploration over the last  decade and contradictory claims have been
  made  (Hartman  2001;  Burningham   et  al.   2005;  Jeffries  2007;
  Hillenbrand et al. 2008 and references their in). In near future, we
  expect to  make further progress in  our study on  this subject by
  enlarging the  size of our  sample, identifying the binaries  in our
  FOV  as  well  as  by  a better  measurement  of  the  intra-cluster
  extinction correction.

Our multi-epoch  observations have  allowed us to  identify in  a very
small  FOV a  large number  periodic variables  (an  additional 50\%).
These newly identified periodic  variables were unidentified mostly by
previous surveys  carried out over  a single observing season.   If we
exclude massive stars or stars  highly affected by nebulosity, then we
found about 44\% of stars in our FoV to be periodic.  This is indeed a
great  gain  with respect  to  previous surveys  of  S99  and H02  who
discovered periodic  variability only in 11\%  and 26\%, respectively,
of stars in our same FOV.  Here  we must keep in mind that the S99 and
H00 surveys  were carried out  with much smaller  aperture telescopes,
hence  on  a  smaller  sample  of brighter  stars.   We  could  obtain
agreement with  previously determined periods in  $\sim$70\% of stars,
could explain the disagreement for 16\% of periodic stars, whereas did
not detect periodicity in the remaining 14\% stars common in all these
surveys.   Our finding, and  the results  of other  ongoing multi-year
surveys  (e.g.  H00)  as  well,   pose  a  strong  warning  about  the
completeness  of  the  periodic  variables discovered  in  young  open
clusters from  single observing run,  which is the most  common case.
If we extend our finding to other star forming regions photometrically
monitored  in only  one season,  we expect  that the  total  number of
periodic variables is highly underestimated.  Therefore, when deriving
any  conclusion about  the  distribution of  rotation  period and  its
implication to  angular momentum evolution,  the sample incompleteness
must be seriously taken into account.

 Our monitoring program together with two previous surveys has yielded
 148 periodic variables in a sample  of 346 stars detected in our FOV.
 Except few stars,  which show no definite variability,  all the other
 stars  were  found  to  be  variables, although  with  no  systematic
 variations, i.e.   irregular variables.   It would be  interesting to
 know what makes  some stars periodic and others  non-periodic. Is the
 undetected  periodicity  due   to  unfavourable  inclination  of  the
 rotation axis, which does not allow the spot pattern visibility to be
 modulated?  Or, is  a very unstable spot pattern  the major source of
 irregular behaviours of these non-periodic variables?  In addition to
 these causes,  sporadic accretion and extinction due  to clumpy inner
 disk  could  be  the  source  of irregular  variation  in  the  stars
 classified as a  CTTS.  We will be addressing  all these questions in
 our forthcoming papers.

\section{Conclusions}

Five consecutive years of observation of a 10$\times$10 arc-minutes region  of  the ONC, using moderate size telescopes has allowed us to find several  interesting results. We summarize our results below:

\begin{itemize}
\item We identified 13 new stars with moderate to strong H$\alpha$ emission, whereas, we did not detect emission in 7 stars, which were reported in the literature as strong H$\alpha$  emitters.
\item The median multi-band magnitudes obtained from the  long-term  time series data have allowed us to  put stellar members on HR diagrams more accurately than the previous studies. From the  comparison with theoretical isochrones, it appears that the age of ONC is 1-2 Myr with a noticeable spread in the age.
\item  Our multi-year monitoring has allowed us to detect 56 new periodic variables, which  increases by 50\% the number  of known periodic variables in the ONC region under study.
\item From a  comparison with previously know periodic variables we found  matching periods for 70\% variables. We find
disagreement for 16\% which can be explained either by period doubling or aliasing. To remaining 14\%  stars no period variation was  detected. 

\item We also find  the absence of stars with period shorter  than 0.9 days  in the 1~Myr old,  brights members ( 12$<$I$<$17 mag) of the ONC.

\item Despite very high background emission, a large number of periodic variables \rm have been identified  inside the nebula. These stars display a higher level of variability as for  as reduced  chi-square ($\chi^2_{\nu}$) is concern.

\item Our study reveals that   about   72\%  of   CTTS   in our FOV are  periodic, whereas,  percentage of periodic WTTS is just  32\%. This indicate that  inhomogeneity patterns on the surface of CTTS  of the ONC are   much more  stable than in WTTS. 
\item All but two stars in our small FOV are found to be variable, according to the chi-square criterion.

\end{itemize}

\section*{Acknowledgements}

The  Observations  reported in  this  paper  were  obtained using  two
telescopes of the Indian Institute of Astrophysics (Bangalore, India):
the 2-m  HCT of Indian  Astronomical Observatory (IAO), and  the 2.3-m
VBT of Vainu Bappu Observatory  (VBO) in Kavalur.  We thank the staffs
of  IAO and  VBO for  their active  support during  the course  of our
observations.  We are  also very much grateful to  Prof. W. Herbst
and Dr.  K.G. Stassun who provided  the raw data  of their photometric
observations. This  work has been also  supported by a  grant given by
the Department of Science \&  Technology India and by the Italian MIUR
(Ministero  dell'Istruzione, Universit\'a  e Ricerca).   The extensive
use  of the  SIMBAD  and ADS  databases  operated by  the CDS  center,
Strasbourg, France, are gratefully acknowledged. Finally, we  are indebted   to  our reviewer Prof. Herbst  for  giving
 valuable suggestion/comments which immensely helped  us to improve the content
 of the  manuscript.

{}

\begin{figure*}
\begin{minipage}{18cm}
\centerline{
\psfig{file=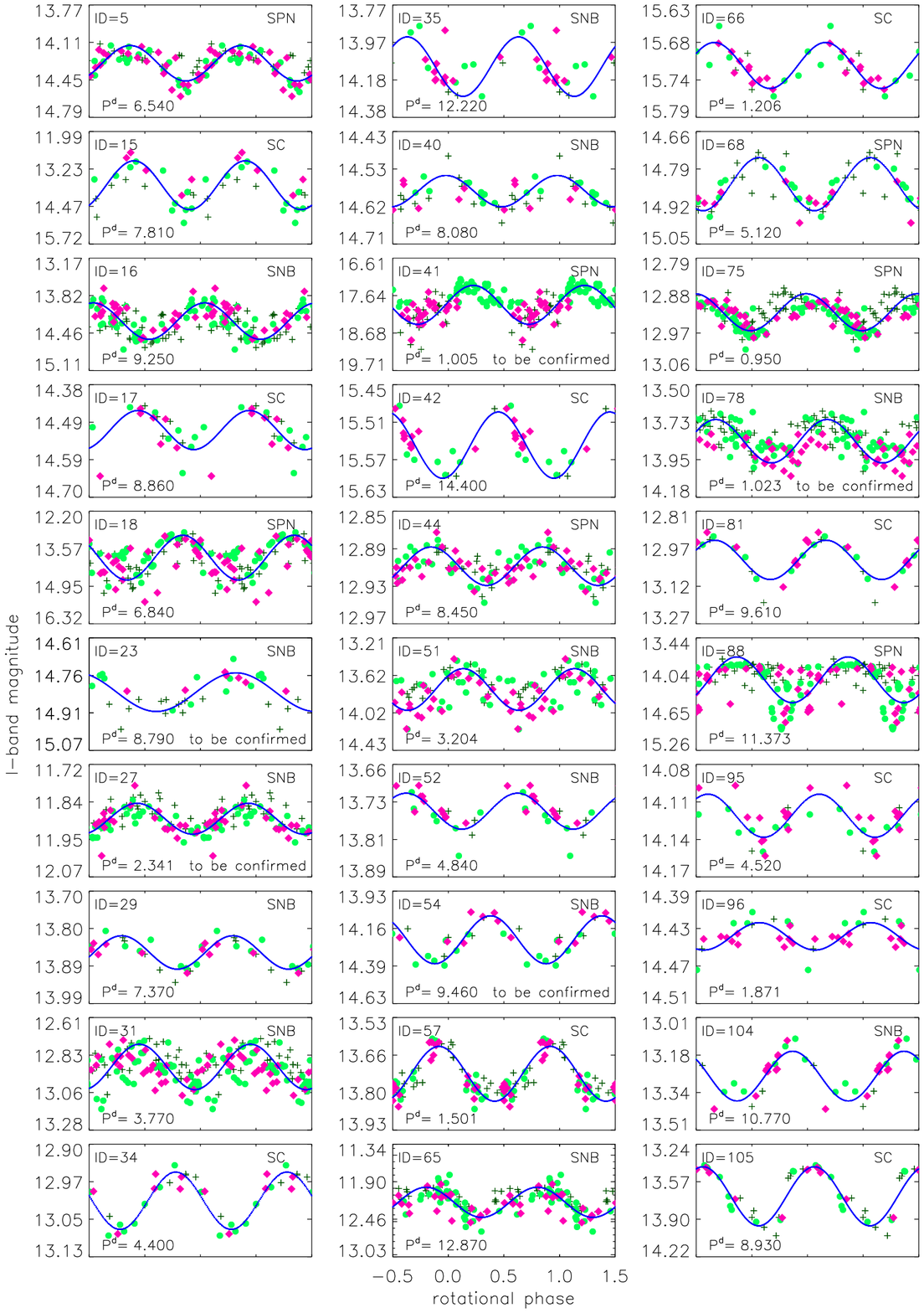}
}
\end{minipage}
\vspace{-2cm}
\caption{I-band light curves of our periodic variables vs. rotation phase. Phases are computed using the rotation period reported in Table\,\ref{tab_results}. Different symbols are used to disctinguish data belonging to different time intervals (see caption of Fig.\,\ref{example_lc}. The solid line represents the sinusoidal fit to the data.}
\label{curve1}
\end{figure*}

\begin{figure*}
\begin{minipage}{18cm}
\centerline{
\psfig{file=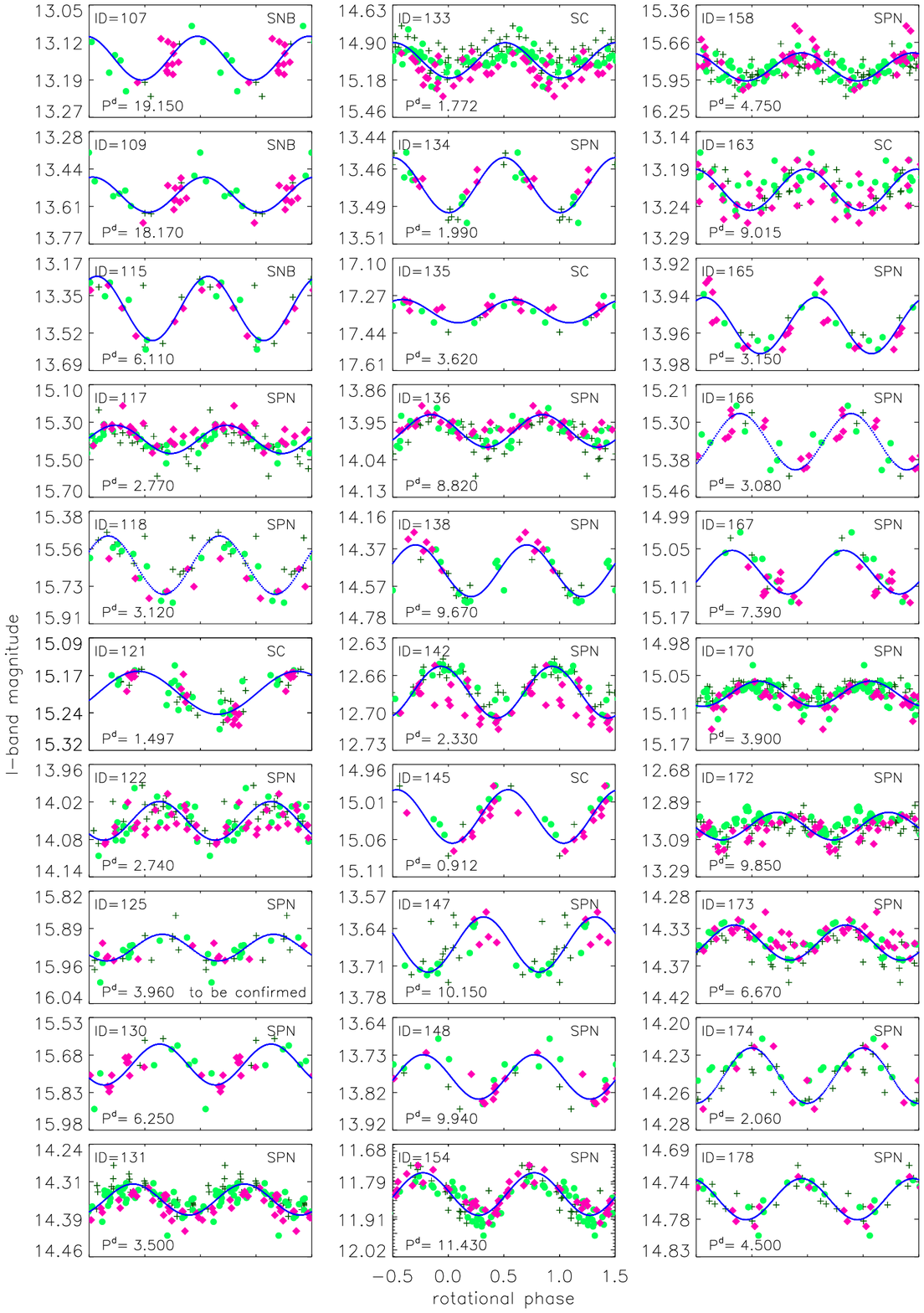}
}
\end{minipage}
\caption{continued.}
\label{curve2}
\end{figure*}

\begin{figure*}
\begin{minipage}{18cm}
\centerline{
\psfig{file=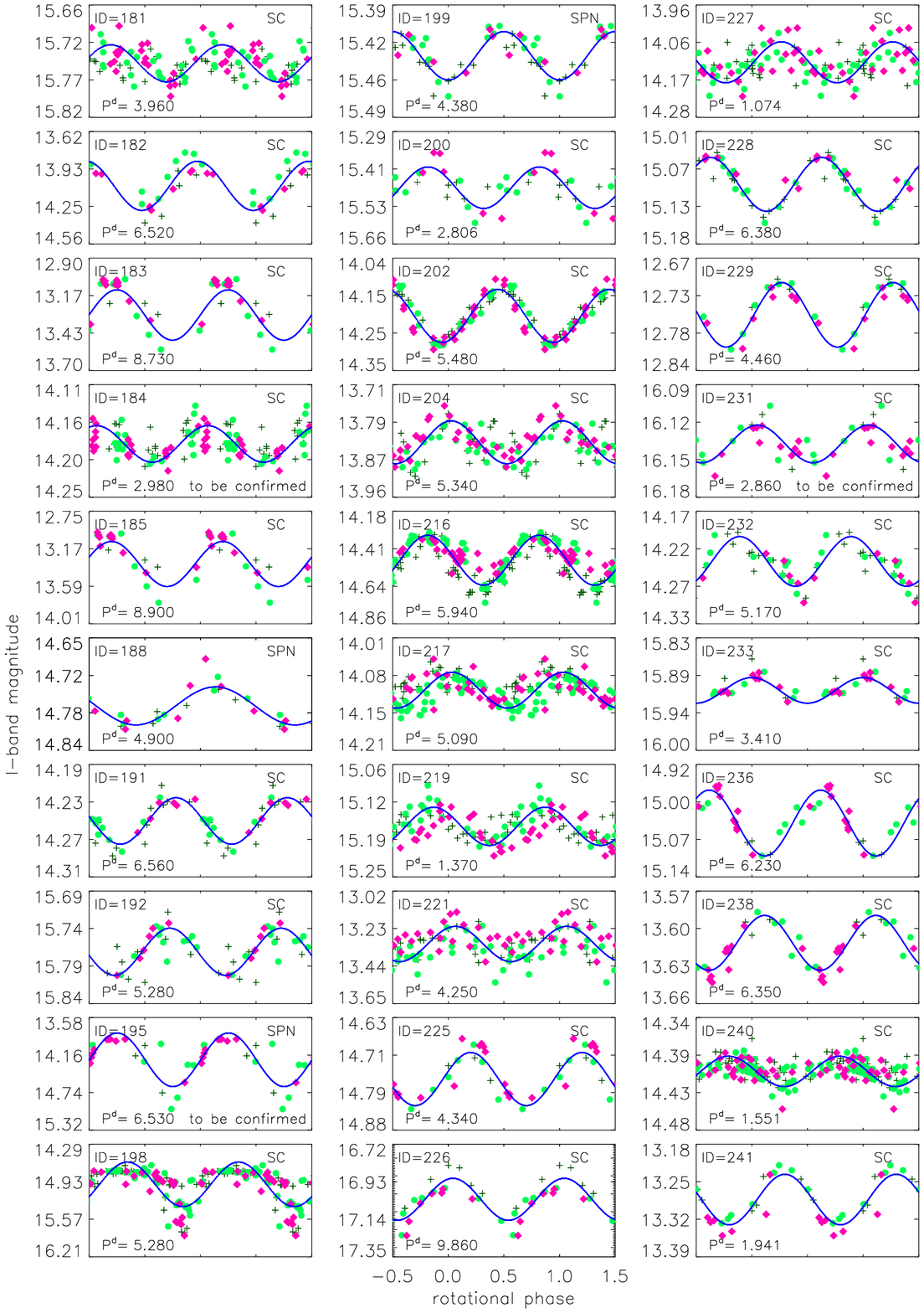}
}
\end{minipage}
\caption{continued}
\label{curve3}
\end{figure*}

\begin{figure*}
\begin{minipage}{18cm}
\centerline{
\psfig{file=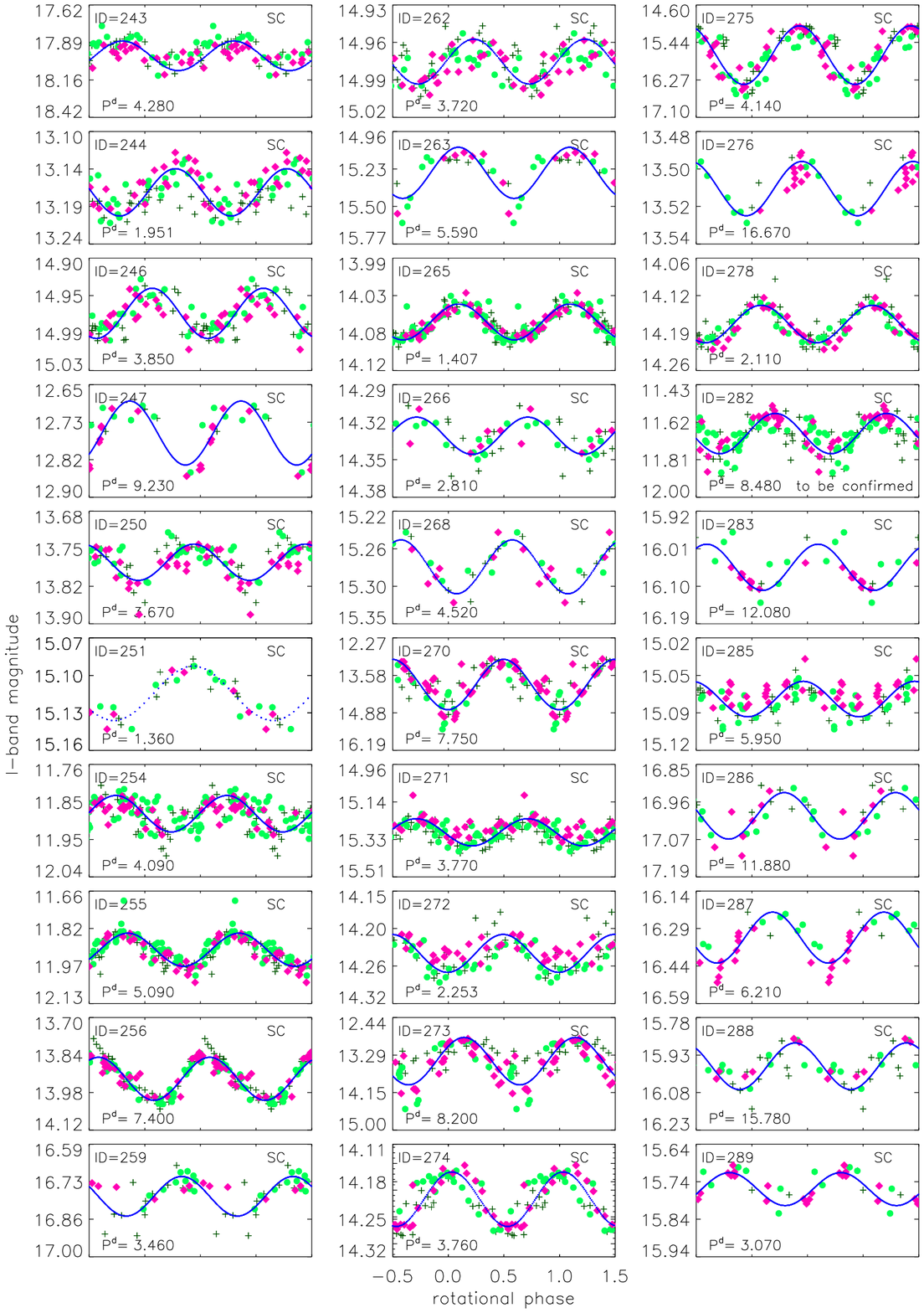}
}
\end{minipage}
\caption{continued}
\label{curve4}
\end{figure*}

\begin{figure*}
\begin{minipage}{18cm}
\centerline{
\psfig{file=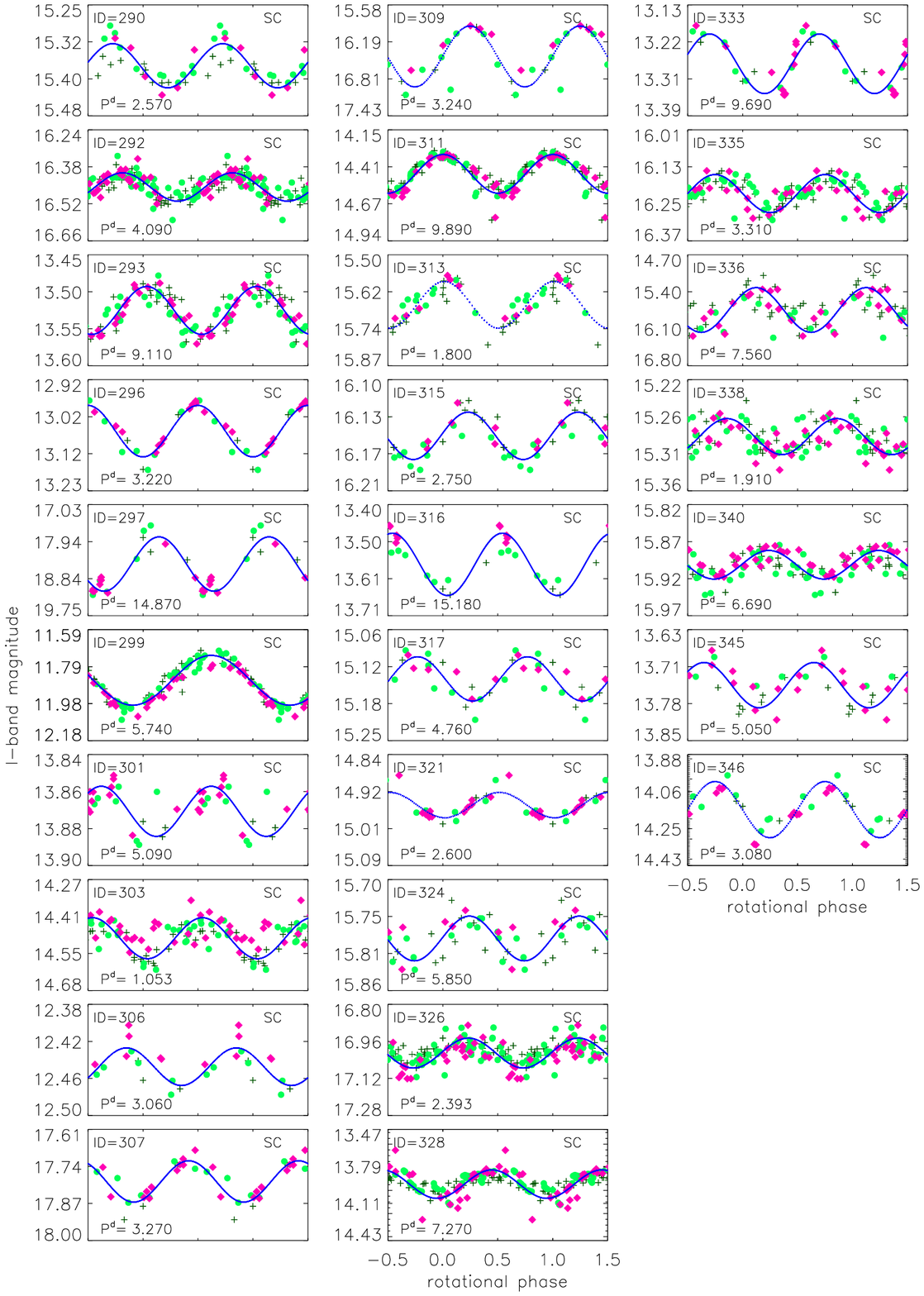}
}
\end{minipage}
\caption{continued}
\label{curve5}
\end{figure*}

\scriptsize
\begin{longtable}{|l|r|r|r|r|@{\hspace{.01cm}}c|r|@{\hspace{.01cm}}r|@{\hspace{.01cm}}c|@{\hspace{.01cm}}c|@{\hspace{.01cm}}l|@{\hspace{.01cm}}l|@{\hspace{.01cm}}c|@{\hspace{.01cm}}c|@{\hspace{.01cm}}c|c|@{\hspace{.01cm}}}
\caption{Result of the periodogram analysis of periodic variables of the ONC.}\label{tab_results}\\
\hline
ID & JW  &  \multicolumn{1}{c|}{Power}  &  \multicolumn{1}{c|}{P$\pm\Delta$P}     &   \multicolumn{1}{c|}{ $\chi^2_{\nu}$}  &  \multicolumn{1}{c|}{$<\sigma>$}  &  \multicolumn{1}{c|}{$\Delta$I}   &  \multicolumn{1}{c|}{\#} &  \multicolumn{1}{c|}{\#} &  \multicolumn{1}{c|}{\#}  &  \multicolumn{1}{c|}{n1}  &   \multicolumn{1}{c|}{n2}  &   \multicolumn{1}{c|}{n3}  &   \multicolumn{1}{c|}{Sky} &   \multicolumn{1}{c|}{Object Type}&   \multicolumn{1}{c|}{Neighbour}  \\

 &     & \multicolumn{1}{c|}{}         &   \multicolumn{1}{c|}{(d)}    &  \multicolumn{1}{c|}{}  & 
 \multicolumn{1}{c|}{(mag)}  &  \multicolumn{1}{c|}{(mag)} &  \multicolumn{1}{c|}{obs.} &  \multicolumn{1}{c|}{mean} &  \multicolumn{1}{c|}{dis.} &  \multicolumn{1}{c|}{}  &  \multicolumn{1}{c|}{} &  \multicolumn{1}{c|}{}  & &  &\\
 \hline 
\endfirsthead
\caption{Continued. }\\
\hline
ID  & JW   &    \multicolumn{1}{c|}{Power}  &  \multicolumn{1}{c|}{P$\pm\Delta$P}     &   \multicolumn{1}{c|}{ $\chi^2_{\nu}$}  &  \multicolumn{1}{c|}{$<\sigma>$}  &  \multicolumn{1}{c|}{$\Delta$I}   &  \multicolumn{1}{c|}{\#} &  \multicolumn{1}{c|}{\#} &  \multicolumn{1}{c|}{\#}  &  \multicolumn{1}{c|}{n1}  &   \multicolumn{1}{c|}{n2}  &   \multicolumn{1}{c|}{n3}  &   \multicolumn{1}{c|}{Sky} &   \multicolumn{1}{c|}{Object Type}&   \multicolumn{1}{c|}{Neighbour} \\
 &   & \multicolumn{1}{c|}{}          &   \multicolumn{1}{c|}{(d)}    &  \multicolumn{1}{c|}{}  & 
 \multicolumn{1}{c|}{(mag)}  &  \multicolumn{1}{c|}{(mag)} &  \multicolumn{1}{c|}{obs.} &  \multicolumn{1}{c|}{means} &  \multicolumn{1}{c|}{dis.} &  \multicolumn{1}{c|}{}  &  \multicolumn{1}{c|}{} &  \multicolumn{1}{c|}{}  & &  &\\
\hline
\endhead

  5  &265   &          45.59  &    6.540 $\pm$  0.100 &  73.17 & 0.041  &    0.34  & 230  &  61  &   2  &   c4  &        c3/c5/H  &     new &   SPN &  C &   y \\
 15  &710   &             15.68  &    7.810 $\pm$  0.180 &    106.99   &  0.231  &    1.43  &  76  &  25  &   2  &   c2  &        c1/H  &      =S &    SC &  C &   - \\
 16  &349   &            50.41  &    9.250 $\pm$  0.120  &    173.62   &  0.054  &    0.63  & 397  & 101  &   6  &   c5  &          c3/c4  &     new &   SNB &  - &   - \\
 17  &125   &              16.71  &    8.860 $\pm$  0.150  &     86.95   &  0.010  &    0.11  &  74  &  21  &   2  &    c3  &          c1/c4  &     new &    SC &  C &   - \\
 18  &278   &             60.60  &    6.840 $\pm$  0.060  &   4189.88   &  0.016  &    1.60  & 400  & 102  &   6  &   c5  &            all  &      =H &   SPN &  C &   - \\
 23$^{\rm a}$  &366   &            18.86  &    8.790 $\pm$  0.230  &     36.30   &  0.026  &    0.16  &  97  &  29  &   1  &   c2  &              -  &     =H &   SNB &  - &   - \\
 27  &437   &             31.38  &    2.341 $\pm$  0.008  &     11.60   &  0.018  &    0.09  & 180  &  88  &   5  &   c5  &              H  &   =H   &   SNB &  C &   - \\
 29  &417   &           15.58  &    7.370 $\pm$  0.100  &      7.64   &  0.035  &    0.08  &  75  &  21  &   1  &   c3  &             c1/H  &     =H &   SNB &  C &   - \\
 31  &622   &             19.93  &    3.770 $\pm$  0.150   &     52.29   &  0.032  &    0.26  & 365  & 105  &   4  &   c5  &           c1/H  &     new &   SNB &  - &   - \\
 34  &81    &           25.05  &    4.400 $\pm$  0.040  &     26.53   &  0.010  &    0.12  &  62  &  21  &   2  &   c4  &     c1/c2/c3/H  &      =H &    SC &  W &   - \\
 35  &9213  &           21.64  &   12.220 $\pm$  1.410  &     16.13   &  0.039  &    0.34  & 103  &  21  &   1  &   c1  &             c3  &     new &   SNB &  - &   y \\
 40  &317   &            16.33  &    8.080 $\pm$  0.190  &     14.37   &  0.018  &    0.08  &  97  &  29  &   3  &   c5  &          c2  &      =H &   SNB &  - &   - \\
 41$^{\rm b}$  &3084  &          79.72  &    1.005 $\pm$  0.001  &     93.87   &  0.165  &    1.07  & 410  & 106  &   8  &   c5  &    c1/c2/c3/c5  &     new &   SPN &  - &   - \\
 42  &71    &          28.67  &   14.400 $\pm$  2.400  &     16.90   &  0.013  &    0.11  & 103  &  21  &   1  &   c1  &        c4/c5/H  &     new &    SC &  - &   - \\
 44  &291   &            30.76  &    8.450 $\pm$  0.210  &      5.93   &  0.009  &    0.04  & 223  &  61  &   2  &    c2  &          c4  &  $\ne$H &   SPN &  - &   - \\
 51  &658   &           28.97  &    3.204 $\pm$  0.004    &  121.67   &  0.090  &    0.45  & 238  &  61  &   3  &  c4  &       c2/c3/c5  &     new &   SNB &  - &   - \\
 52  &356   &              27.28  &    4.840 $\pm$  0.260  &     45.21   &  0.008  &    0.08  & 103  &  21  &   1  &    c1  &        c3/c4/c5/H  &      =H &   SNB &  - &   - \\
 54$^{\rm a}$  &-     &            15.30  &    9.460 $\pm$  0.170  &     45.33   &  0.072  &    0.30  &  77  &  21  &   2  &    c3  &              -  &     new &   SNB &  - &   - \\
 57  &77    &             80.10  &    1.501 $\pm$  0.004  &     47.05   &  0.013  &    0.19  & 245  &  62  &   2  &    c3  &            all  &     =HS &    SC &  W &   - \\
 65  &678   &           31.20  &   12.870 $\pm$  0.240  &    120.88   &  0.073  &    0.45  & 203  &  93  &   3  &  c5  &        c4/H  &      =H &   SNB &  W &   - \\
 66  &130   &            22.84  &    1.206 $\pm$  0.002  &     12.60   &  0.027  &    0.10  & 417  & 107  &   5  &    c5  &            all  &      =H &    SC &  - &   - \\
 68  &292   &             25.68  &    5.120 $\pm$  0.070   &     31.56   &  0.019  &    0.18  &  97  &  29  &   3  &    c2  &     c1/c4/c5/H  &     new &   SPN &  - &   - \\
 75  &275   &            80.36  &    0.950 $\pm$  0.001   &     18.84   &  0.014  &    0.09  & 411  & 107  &   6  &   c5  &     c1/c2/c4/H  &      =H &   SPN &  - &   - \\
 78$^{\rm b}$  &516   &            51.65  &    1.023 $\pm$  0.001 &     45.39   &  0.035  &    0.26  & 415  & 107  &   4  &    c5  &             c1  &     new &   SNB &  - &   - \\
 81  &123   &              37.80  &    9.610 $\pm$  1.050  &    330.29   &  0.004  &    0.16  &  90  &  21  &   1  &   c1  &            all  &  $\ne$S &    SC &  - &   - \\
 88  &323   &           54.49  &   11.370 $\pm$  0.020  &    766.12   &  0.021  &    0.73  & 417  & 107  &   5  &   c5  &       c1/c2/c3  &     new &   SPN &  C &   - \\
 95  &152   &             23.55  &    4.520 $\pm$  0.240  &      9.61   &  0.006  &    0.03  & 103  &  21  &   1  &   c1  &            all  &     new &    SC &  - &   - \\
 96  &183   &              25.00  &    1.871 $\pm$  0.005  &      9.05   &  0.007  &    0.03  & 103  &  21  &   1  &   c1  &        c3/c4/H  &  $\ne$H &    SC &  - &   - \\
104  &636   &             36.26  &   10.770 $\pm$  1.320  &    133.92   &  0.012  &    0.22  & 101  &  21  &   1  &   c1  &        c2/c4/H  &      =H &   SNB &  C &   y \\
105  &192   &              36.08  &    8.930 $\pm$  0.230  &   1090.97   &  0.008  &    0.51  &  97  &  29  &   2  &    c2  &            all  &      =S &    SC &  C &   - \\
107  &551   &           32.64  &   19.150 $\pm$  4.190  &     24.69   &  0.010  &    0.09  & 103  &  21  &   1  &   c1  &             c5  &     new &   SNB &  - &   y \\
109  &535   &           36.81  &   18.170 $\pm$  3.780  &     33.88   &  0.014  &    0.15  & 103  &  21  &   1  &   c1  &             c5  &     new &   SNB &  - &   - \\
115  &674   &             22.80  &    6.110 $\pm$  0.070  &    139.90   &  0.014  &    0.30  &  62  &  20  &   2  &   c3  &       c1/c2/c4  &     new &   SNB &  - &   - \\
117  &299   &              28.69  &    2.770 $\pm$  0.010  &     45.08   &  0.025  &    0.15  & 245  &  62  &   3  &   c4  &             c1  &      =H &   SPN &  C &   - \\
118  &3104  &            49.41  &    3.120 $\pm$  0.020  &     55.95   &  0.033  &    0.28  &  97  &  29  &   2  &   c2  &            all  &      =H &   SPN &  - &   - \\
121  &70    &               45.74  &    1.497 $\pm$  0.005  &     19.48   &  0.011  &    0.09  & 246  &  62  &   4  &   c4  &        c1/c2/H  &      =H &    SC &  - &   - \\
122  &250   &              30.21  &    2.740 $\pm$  0.010   &      9.13   &  0.012  &    0.06  & 245  &  62  &   1  &   c4  &            all  &      =H &   SPN &  - &   - \\
125$^{\rm a}$  &234   &            16.92  &    3.960 $\pm$  0.040  &      7.01   &  0.016  &    0.05  &  97  &  29  &   2  &    c2  &              -  &     new &   SPN &  - &   - \\
130  &324   &              27.69  &    6.250 $\pm$  0.440  &     43.88   &  0.016  &    0.18  & 103  &  21  &   2  &   c1  &          c4/c5  &     new &   SPN &  C &   - \\
131  &254   &              21.10  &    3.500 $\pm$  0.010  &     11.30   &  0.017  &    0.06  & 417  & 107  &   5  &    c5  &        c1/c2/H  &      =H &   SPN &  - &   - \\
133  &104   &             58.44  &    1.772 $\pm$  0.004   &    161.70   &  0.015  &    0.26  & 417  & 107  &   5  &   c5  &        c3/c4/H  &      =H &    SC &  C &   - \\
134  &280   &            31.71  &    1.990 $\pm$  0.010  &     10.55   &  0.008  &    0.04  &  96  &  29  &   2  &    c2  &            all  &     =HS &   SPN &  - &   - \\
135  &-     &             19.17  &    3.620 $\pm$  0.140  &      4.83   &  0.031  &    0.11  & 103  &  21  &   1  &    c1  &              H  &  $\ne$H &    SC &  - &   - \\
136  &245   &              43.45  &    8.820 $\pm$  0.190  &     43.95   &  0.006  &    0.09  & 245  &  62  &   2  &  c4  &          c2/c5  &      =S &   SPN &  C &   - \\
138  &485   &            36.25  &    9.670 $\pm$  0.270  &    205.72   &  0.013  &    0.29  &  97  &  29  &   2  &    c2  &            all  &      =H &   SPN &  - &   - \\
142  &220   &              52.08  &    2.330 $\pm$  0.010  &     15.02   &  0.005  &    0.05  & 208  &  58  &   3  &   c4  &        c1/H  &      =H &   SPN &  C &   - \\
145  &168   &             34.51  &    0.912 $\pm$  0.009 &     16.10   &  0.010  &    0.07  & 103  &  21  &   1  &   c1  &        c4/c5/H  &      =H &    SC &  W &   - \\
147  &243   &            24.49  &   10.150 $\pm$  0.300  &     52.89   &  0.006  &    0.10  &  97  &  29  &   1  &    c2  &            all  &      =S &   SPN &  C &   - \\
148  &258   &             31.36  &    9.940 $\pm$  1.130  &     76.22   &  0.006  &    0.11  & 103  &  21  &   1  &    c1  &        c4/c5/H  &      =H &   SPN &  C &   - \\
154  &683   &            17.75  &   11.430 $\pm$  0.190  &     12.92   &  0.028  &    0.13  & 205  &  91  &   2  &   c5  &        c4/H  &      =H &   SPN &  - &   y \\
158  &573   &            55.73  &    4.750 $\pm$  0.030   &     14.14   &  0.043  &    0.23  & 417  & 107  &   4  &  c5  &             c3  &     new &   SPN &  C &   - \\
163  &117   &             37.58  &    9.010 $\pm$  0.500  &     10.08   &  0.010  &    0.06  & 244  &  62  &   2  &    c4  &             c1  &  =H &    SC &  C &   - \\
165  &281   &             29.75  &    3.150 $\pm$  0.110   &      6.15   &  0.005  &    0.03  & 103  &  21  &   1  &   c1  &           c4/H  &      =H &   SPN &  - &   - \\
166  &290   &             31.16  &    3.080 $\pm$  0.010  &     45.72   &  0.009  &    0.12  & 103  &  21  &   2  &    c1  &          c3  &     new &   SPN &  - &   - \\
167  &283   &            35.35  &    7.390 $\pm$  0.620   &     27.85   &  0.009  &    0.07  & 103  &  21  &   2  &   c1  &        c4/c5/H  &      =H &   SPN &  - &   - \\
170  &392   &             16.05  &    3.900 $\pm$  0.020 &      3.07   &  0.015  &    0.04  & 416  & 107  &   5  &    c3  &             c5  &     new &   SPN &  C &   - \\
172  &268   &             39.08  &    9.850 $\pm$  0.140  &     69.50   &  0.016  &    0.15  & 417  & 107  &   4  &   c5  &     c1/c3/c4/H  &      =H &   SPN &  C &   - \\
173  &477   &              39.36  &    6.670 $\pm$  0.110  &     10.25   &  0.007  &    0.04  & 245  &  62  &   3  &  c4  &       c1/c3  &     new &   SPN &  - &   - \\
174  &415   &              16.75  &    2.060 $\pm$  0.010  &     10.25   &  0.007  &    0.04  & 245  &  62  &   3  &  c2  &          c1  &     new &   SPN &  - &   - \\
178  &427   &              24.86  &    4.500 $\pm$  0.060  &     15.82   &  0.008  &    0.05  &  97  &  29  &   2  &    c2  &          c1  &     new &   SPN &  C &   - \\
181  &144   &              27.78  &    3.960 $\pm$  0.030 &      9.23   &  0.012  &    0.05  & 245  &  62  &   2  &    c4  &          c1/c3  &  $\ne$H &    SC &  - &   - \\
182  &164   &              25.28  &    6.520 $\pm$  0.080  &    643.74   &  0.018  &    0.41  &  77  &  21  &   2  &    c3  &            all  &      =H &    SC &  - &   - \\
183  &137   &              29.62  &    8.730 $\pm$  0.870  &   1027.81   &  0.009  &    0.36  & 103  &  21  &   2  &   c1  &            all  &     =H &    SC &  - &   y \\
184$^{\rm a}$  &272   &             45.55  &    2.980 $\pm$  0.020   &      7.05   &  0.008  &    0.05  & 245  &  62  &   3  &   c4  &              -  &     new &    SC &  C &   - \\
185  &136   &              32.00  &    8.900 $\pm$  0.900  &   1062.74   &  0.009  &    0.51  & 103  &  21  &   1  &  c1  &            all  &      =H &    SC &  C &   y \\
188  &418   &             25.56  &    4.900 $\pm$  0.040   &     12.98   &  0.013  &    0.06  &  77  &  21  &   2  &   c3  &            all  &     new &   SPN &  - &   - \\
191  &188   &              37.52  &    6.560 $\pm$  0.120  &     12.49   &  0.008  &    0.05  &  97  &  29  &   2  &  c2  &            all  &     =HS &    SC &  - &   - \\
192  &105   &              22.84  &    5.280 $\pm$  0.080   &     15.20   &  0.009  &    0.07  &  97  &  29  &   2  &    c2  &             c1  &     new &    SC &  - &   - \\
195  &672   &               32.09  &    6.530 $\pm$  0.110  &   1538.41   &  0.009  &    0.83  & 103  &  21  &   1  &   c1  &              -  &     =HS &   SPN &  C &   - \\
198  &102   &             111.78  &    5.280 $\pm$  0.040  &   1369.67   &  0.016  &    0.76  & 417  & 107  &   7  &   c5  &          c3/c4  &     new &    SC &  C &   - \\
199  &486   &              25.52  &    4.380 $\pm$  0.050 &      5.72   &  0.010  &    0.05  &  97  &  29  &   1  &    c2  &          c1/c3  &     new &   SPN &  W &   - \\
200  &149   &              20.15  &    2.806 $\pm$  0.003   &     44.96   &  0.014  &    0.14  &  77  &  21  &   2  &   c3  &              H  &      =S &    SC &  C &   - \\
201  &704   &              28.59  &    6.790 $\pm$  0.030   &     34.58   &  0.026  &    0.16  & 243  &  62  &   5  &   c4  &           c1/H  &     new &   SPN &  - &   y \\
202  &211   &               100.77  &    5.480 $\pm$  0.050  &    129.97   &  0.005  &    0.15  & 245  &  62  &   1  &    c4  &            all  &     =HS &    SC &  C &   - \\
204  &334   &              41.31  &    5.340 $\pm$  0.070  &     37.93   &  0.009  &    0.10  & 245  &  62  &   2  &  c4  &             c1  &      =S &    SC &  C &   - \\
216  &422   &             142.43  &    5.940 $\pm$  0.050  &    330.26   &  0.012  &    0.30  & 417  & 107  &   5  &   c5  &       c1/c3/c4  &      =S &    SC &  C &   - \\
217  &591   &              44.60  &    5.090 $\pm$  0.030  &      9.42   &  0.014  &    0.06  & 417  & 107  &   9  &    c5  &             c1  &     new &    SC &  - &   - \\
219  &148   &                30.60  &    1.370 $\pm$  0.010   &     18.31   &  0.010  &    0.07  & 245  &  62  &   1  &  c4  &     c2/c3/c5/H  &     =HS &    SC &  C &   - \\
221  &502   &             31.09  &    4.250 $\pm$  0.010   &    206.77   &  0.008  &    0.19  & 245  &  62  &   3  &    c4  &             c1  &     new &    SC &  - &   - \\
225  &138   &              40.46  &    4.340 $\pm$  0.210  &     70.62   &  0.008  &    0.12  & 103  &  21  &   1  &  c1  &           c5/H  &      =S &    SC &  C &   - \\
226  &543   &             25.50  &    9.860 $\pm$  0.190 &     75.04   &  0.029  &    0.23  &  77  &  21  &   2  &    c3  &    c1/c2/c4/c5  &     new &    SC &  - &   - \\
227  &107   &             28.63  &    1.074 $\pm$  0.005  &     48.85   &  0.009  &    0.12  & 245  &  62  &   3  &   c4  &          c1/c3  &     new &    SC &  C &   - \\
228  &326   &              34.15  &    6.380 $\pm$  0.120  &     17.71   &  0.010  &    0.08  &  97  &  29  &   2  &   c2  &            all  &      =H &    SC &  - &   - \\
229  &239   &             28.44  &    4.460 $\pm$  0.210  &     60.67   &  0.004  &    0.10  &  76  &  18  &   1  &    c1  &     c3/c4/c5/H  &     =HS &    SC &  C &   - \\
231$^{\rm a}$  &446   &           17.18  &    2.860 $\pm$  0.090  &      3.15   &  0.012  &    0.03  & 103  &  21  &   1  &     c1  &              -  &     new &    SC &  - &   - \\
232  &222   &               25.53  &    5.170 $\pm$  0.070  &     18.00   &  0.009  &    0.08  &  97  &  29  &   2  &   c2  &            all  &     =HS &    SC &  C &   - \\
233  &318   &               24.80  &    3.410 $\pm$  0.030  &      5.96   &  0.010  &    0.04  & 103  &  21  &   1  &    c1  &        c3/c4/H  &      =H &    SC &  - &   - \\
236  &311   &             41.23  &    6.230 $\pm$  0.440  &     65.69   &  0.006  &    0.13  & 103  &  21  &   2  &    c1  &            all  &     =HS &    SC &  - &   - \\
238  &76    &              60.99  &    6.350 $\pm$  0.110  &     10.12   &  0.012  &    0.06  & 245  &  62  &   2  &     c4  &            all  &      =H &    SC &  - &   - \\
240  &120   &                20.10  &    1.551 $\pm$  0.003 &      2.88   &  0.012  &    0.04  & 417  & 107  &   5  &   c5  &             c4  &  $\ne$S &    SC &  C &   - \\
241  &158   &               18.81  &    1.941 $\pm$  0.007   &     49.35   &  0.006  &    0.09  &  60  &  21  &   2  &   c3  &            all  &      =S &    SC &  - &   - \\
243  & -     &               25.08  &    4.280 $\pm$  0.060   &      2.81   &  0.087  &    0.19  & 245  &  62  &   2  &   c4  &       c1/c4/c5  &     new &    SC &  - &   - \\
244  &576   &                57.83  &    1.951 $\pm$  0.009  &      9.61   &  0.010  &    0.06  & 243  &  62  &   3  &    c4  &          c2/c5/H  &     =H &    SC &  C &   - \\
246  &155   &               50.63  &    3.850 $\pm$  0.040   &      8.50   &  0.010  &    0.05  & 245  &  62  &   3  &    c4  &    c1/c3/c4/c5  &     new &    SC &  C &   - \\
247  &252   &              22.89  &    9.230 $\pm$  0.930  &    115.14   &  0.005  &    0.14  &  80  &  21  &   1  &    c1  &              H  &     new &    SC &  C &   y \\
250  &135   &                50.30  &    3.670 $\pm$  0.030   &     29.59   &  0.009  &    0.08  & 245  &  62  &   1  &   c4  &        c1/c5/H  &      =H &    SC &  C &   - \\
251  &181   &               26.76  &    1.360 $\pm$  0.005  &      9.16   &  0.008  &    0.04  &  97  &  29  &   2  &   c2  &            all  &     =HS &    SC &  C &   - \\
254  &330   &              15.27  &    4.090 $\pm$  0.060   &     10.99   &  0.031  &    0.09  & 386  & 105  &   1  &   c5  &          c4/H  &     $\ne$H &    SC &  W &   - \\
255  &232   &              72.27  &    5.090 $\pm$  0.030  &     37.90   &  0.021  &    0.14  & 374  & 107  &   3  &    c5  &          c4/H  &     =H &    SC &  W &   - \\
256  &65    &              135.50  &    7.400 $\pm$  0.080   &     79.32   &  0.013  &    0.17  & 366  & 101  &   5  &   c5  &            all  &      =S &    SC &  W &   - \\
259  &3130  &               23.02  &    3.460 $\pm$  0.030   &     39.20   &  0.017  &    0.21  &  97  &  29  &   1  &   c2  &            all  &      =H &    SC &  C &   - \\
262  &218   &               30.82  &    3.720 $\pm$  0.020  &      3.84   &  0.010  &    0.04  & 245  &  62  &   1  &    c4  &            all  &     new &    SC &  - &   - \\
263  &379   &               25.35  &    5.590 $\pm$  0.060  &   1032.75   &  0.008  &    0.37  &  77  &  21  &   1  &   c3  &            all  &      =H &    SC &  C &   - \\
265  &428   &               67.20  &    1.407 $\pm$  0.003   &      6.67   &  0.011  &    0.04  & 417  & 107  &   5  &   c5  &          c3/c4  &     new &    SC &  - &   - \\
266  &556   &              16.18  &    2.810 $\pm$  0.020  &      9.06   &  0.005  &    0.03  &  97  &  29  &   1  &   c2  &          c1/c4  &     new &    SC &  C &   - \\
268  &577   &              20.99  &    4.520 $\pm$  0.040  &     13.73   &  0.014  &    0.06  &  77  &  21  &   2  &   c3  &       c1/c2/c4  &     new &    SC &  C &   - \\
270  &381   &               65.54  &    7.750 $\pm$  0.080  &   8992.91   &  0.016  &    1.76  & 320  &  81  &   5  &    c5  &           c4/H  &      =H &    SC &  C &   - \\
271  &632   &               31.86  &    3.770 $\pm$  0.010  &     35.57   &  0.034  &    0.13  & 413  & 106  &  10  &   c5  &        c1/c4/H  &      =H &    SC &  C &   - \\
272  &628   &                40.58  &    2.253 $\pm$  0.002  &     11.11   &  0.008  &    0.06  & 245  &  62  &   4  &    c4  &           c3/H  &     =HS &    SC &  C &   - \\
273  &647   &               95.82  &    8.200 $\pm$  0.800   &   4771.02   &  0.018  &    1.08  & 314  &  80  &   6  &   c5  &            all  &     new &    SC &  C &   - \\
274  &174   &               80.92  &    3.760 $\pm$  0.030   &     66.39   &  0.007  &    0.11  & 245  &  62  &   5  &   c4  &            all  & $\ne$HS &    SC &  C &   - \\
275  &571   &              80.32  &    4.140 $\pm$  0.050  &   2997.36   &  0.014  &    1.30  & 245  &  62  &   3  &    c4  &            all  &     new &    SC &  C &   - \\
276  &91    &             36.26  &   16.670 $\pm$  3.170 &      5.46   &  0.005  &    0.03  & 103  &  21  &   3  &   c1  &           c5/H  &      =H &    SC &  C &   - \\
278  &416   &              80.91  &    2.110 $\pm$  0.010  &     22.53   &  0.009  &    0.07  & 245  &  62  &   4  &    c4  &            all  &      =S &    SC &  - &   - \\
282$^{\rm a}$  &421   &             32.10  &    8.480 $\pm$  0.100  &     41.38   &  0.029  &    0.20  & 265  & 101  &   3  &   c5  &              -  &     new &    SC &  C &   - \\
283  &3134  &           24.21  &   12.080 $\pm$  1.670   &     15.20   &  0.015  &    0.11  & 103  &  21  &   1  &    c1  &       c2/c3/c4  & $\ne$HS &    SC &  - &   - \\
285  &517   &               20.44  &    5.950 $\pm$  0.020  &      3.16   &  0.009  &    0.03  & 245  &  62  &   1  &   c4  &             c1  &  $\ne$S &    SC &  - &   - \\
286  &5147  &             19.47  &   11.880 $\pm$  0.280   &     40.71   &  0.018  &    0.14  &  77  &  21  &   2  &    c3  &       c1/c2/c4  &     new &    SC &  C &   - \\
287  &633   &              30.60  &    6.210 $\pm$  0.440   &     27.03   &  0.016  &    0.20  & 103  &  21  &   1  &  c1  &          c2/c5  &     new &    SC &  - &   - \\
288$^{\rm a}$  &96      &       22.49  &   15.780 $\pm$  0.740 &     37.07   &  0.014  &    0.19  &  97  &  29  &   1  &   c2  &              -  &     new &    SC &  C &   - \\
289  &200   &               28.21  &    3.070 $\pm$  0.100   &     38.40   &  0.009  &    0.09  & 103  &  21  &   1  &   c1  &            all  &      =H &    SC &  C &   - \\
290  &294   &               16.58  &    2.57 $\pm$  0.070  &     43.74   &  0.007  &    0.09  &  97  &  29  &   1  &   c2  &        c1/c4/H  &      =H &    SC &  C &   - \\
292  &3138  &              24.49  &    4.090 $\pm$  0.020  &     11.74   &  0.033  &    0.11  & 416  & 107  &   6  &   c5  &            all  &      =H &    SC &  - &   - \\
293  &159   &               73.41  &    9.110 $\pm$  0.200 &     10.99   &  0.012  &    0.07  & 245  &  62  &   3  &   c4  &        c3/c5/H  &     $\ne$H &    SC &  - &   - \\
296  &673   &              46.42  &    3.220 $\pm$  0.110  &    311.01   &  0.004  &    0.15  & 101  &  21  &   2  &    c1  &     c2/c4/c5/H  &  $\ne$H &    SC &  C &   - \\
297  &-     &             37.06  &   14.870 $\pm$  2.530  &   1070.23   &  0.100  &    1.37  & 103  &  21  &   3  &   c1  &            all  &      =H &    SC &  - &   - \\
299  &165   &               94.85  &    5.740 $\pm$  0.040  &     98.42   &  0.027  &    0.26  & 281  &  97  &   1  &    c5  &          c4/H  &     =H &    SC &  C &   - \\
301  &380   &               20.30  &    5.090 $\pm$  0.250  &     12.48   &  0.004  &    0.03  & 103  &  21  &   2  &    c1  &        c3/c4/H  &      =H &    SC &  C &   - \\
303  &101   &              36.25  &    1.053 $\pm$  0.004  &    102.95   &  0.008  &    0.15  & 245  &  62  &   4  &    c4  &          c3/c5  &  $\ne$S &    SC &  C &   - \\
306  &375   &               23.34  &    3.060 $\pm$  0.100  &     18.78   &  0.005  &    0.04  &  68  &  15  &   1  &    c1  &             c4  &     new &    SC &  - &   - \\
307  &-     &            29.83  &    3.270 $\pm$  0.120  &      4.91   &  0.038  &    0.14  & 103  &  21  &   1  &   c1  &        c4/c5/H  &  $\ne$H &    SC &  - &   - \\
309  &228   &               25.49  &    3.240 $\pm$  0.020  &    769.84   &  0.045  &    1.01  &  77  &  21  &   3  &    c3  &            all  &      =H &    SC &  - &   - \\
311  &565   &              152.00  &    9.890 $\pm$  0.140  &    292.73   &  0.015  &    0.26  & 416  & 107  &  10  &   c5  &            all  &     new &    SC &  - &   - \\
313  &649   &              36.03  &    1.800 $\pm$  0.009  &    102.33   &  0.007  &    0.15  &  97  &  29  &   2  &   c2  &            all  &      =H &    SC &  C &   - \\
315  &227   &               30.72  &    2.750 $\pm$  0.020  &      8.87   &  0.010  &    0.06  &  97  &  29  &   2  &   c2  &             c1  &     new &    SC &  - &   - \\
316  &728   &             18.90  &   15.180 $\pm$  2.970   &    119.24   &  0.008  &    0.15  &  51  &  15  &   1  &   c1  &             c5  &      =H &    SC &  C &   - \\
317  &94    &               20.27  &    4.760 $\pm$  0.040  &     13.17   &  0.015  &    0.08  &  77  &  21  &   3  &  c3  &            all  &      =H &    SC &  C &   - \\
321  &447   &               34.70  &    2.600 $\pm$  0.070   &     41.10   &  0.006  &    0.06  & 103  &  21  &   2  &    c1  &           c3/H  &      =S &    SC &  C &   - \\
324  &186   &               17.81  &    5.850 $\pm$  0.100   &     15.85   &  0.012  &    0.08  &  97  &  29  &   1  &    c2  &            all  &     =HS &    SC &  C &   - \\
326  &5159  &              30.26  &    2.393 $\pm$  0.008   &     10.34   &  0.040  &    0.13  & 417  & 107  &   6  &   c5  &          c1/c4  &     new &    SC &  C &   - \\
328  &498   &               62.41  &    7.270 $\pm$  0.070    &    217.19   &  0.017  &    0.22  & 417  & 107  &   6  &    c5  &            all  &     =H &    SC &  - &   - \\
333  &501   &              37.24  &    9.690 $\pm$  1.070   &    137.85   &  0.005  &    0.14  & 103  &  21  &   5  &   c1  &            all  &     new &    SC &  C &   - \\
335  &682   &          33.73  &    3.310 $\pm$  0.010  &     13.12   &  0.020  &    0.13  & 282  &  70  &   5  &    c5  &     c1/c3/c4/H  &     new &    SC &  - &   - \\
336  &449   &               45.21  &    7.560 $\pm$  0.140    &   1980.86   &  0.014  &    0.85  & 174  &  43  &   3  &   c4  &             c1  &     new &    SC &  C &   - \\
338  &100   &             18.00  &    1.910 $\pm$  0.030  &      4.79   &  0.013  &    0.04  & 246  &  62  &   4  &    c4  &             c5  & $\sim$H &    SC &  - &   - \\
340  &216   &              25.76  &    6.690 $\pm$  0.110  &      2.42   &  0.014  &    0.04  & 244  &  62  &   2  &   c4  &        c2/c3/H  &     new &    SC &  - &   - \\
345  &639   &              65.53  &    5.050 $\pm$  0.080  &     33.59   &  0.015  &    0.08  &  72  &  21  &   1  &  c4  &            all  &     =HS &    SC &  - &   - \\
346  &284   &             36.68  &    3.080 $\pm$  0.100  &    292.15   &  0.006  &    0.28  & 101  &  20  &   1  &   c1  &        c2/c5/H  &     =HS &    SC &  C &   - \\
\hline
\multicolumn{16}{l}{a: The rotation period was detected in only one season and, although with a  FAP $<$ 1\%, it needs to be confirmed by future observations.}\\
\multicolumn{16}{l}{b: The rotation period, although detected in multiple seasons and with a  FAP $<$ 1\%,  may be  a beat period, being very close to the window function main peak.}
\end{longtable}

\end{document}